\documentclass[11pt, a4paper, onecolumn, copyright, goog]{google}
\usepackage{subcaption}
\usepackage{comment}
\usepackage{multirow}
\usepackage{hyperref}
\hypersetup{
    colorlinks=true,
    linkcolor=blue,
    urlcolor=blue,
}

\usepackage[utf8]{inputenc}
\usepackage{xcolor} % Required for colors

\uselogo{}

\usepackage{dm-colors}
\usepackage{amsmath}
\usepackage{verbatim}
\usepackage{scalerel}
\usepackage{longtable}
\usepackage{array}
\usepackage{booktabs}
\usepackage{enumitem}
\usepackage{xspace}
\usepackage{bm}
\usepackage{mathtools}
\usepackage{soul}
\usepackage[normalem]{ulem}
\usepackage{graphicx}
\usepackage{rotating}
\usepackage{tcolorbox}
\usepackage{listings}
\colorlet{listingbg}{black!5}
\colorlet{listingbreak}{gray}
\colorlet{pykeyword}{dmblue500}
\colorlet{pytype}{dmpurple500}
\colorlet{pystring}{teal!80!black}
\colorlet{pycomment}{dmgray600}
\colorlet{pydecorator}{dmpink500}
\lstdefinestyle{listingbox}{
    basicstyle=\ttfamily\scriptsize,
    breaklines=true,
    breakatwhitespace=false,
    columns=fullflexible,
    keepspaces=true,
    upquote=true,
    showstringspaces=false,
    frame=single,
    rulecolor=\color{listingbg},
    backgroundcolor=\color{listingbg},
    framesep=4pt,
    xleftmargin=4pt,
    xrightmargin=4pt,
    postbreak=\mbox{\textcolor{listingbreak}{$\hookrightarrow$}\space},
}

\lstdefinestyle{prompt}{
    style=listingbox,
    extendedchars=true,
    postbreak={},
    breakindent=0pt,
    literate={“}{\textquotedblleft}1 {”}{\textquotedblright}1 {’}{\textquoteright}1 {—}{\textemdash}1 {–}{\textendash}1 {→}{$\rightarrow$}1,
}
\lstdefinestyle{pyschema}{
    style=listingbox,
    language=Python,
    keywordstyle=\bfseries\color{pykeyword},
    keywordstyle=[2]\color{pytype},
    morekeywords=[2]{str,bool,List,StrEnum},
    stringstyle=\color{pystring},
    commentstyle=\itshape\color{pycomment},
    moredelim=[l][\color{pydecorator}]{@},
}
\usepackage{amssymb}
\usepackage{colortbl}
\usepackage{csquotes}
\usepackage{setspace}
\usepackage{tabularx,ragged2e}
\usepackage{placeins}
\usepackage{afterpage}
\usepackage[symbol]{footmisc}
\usepackage[bibstyle=nature,citestyle=numeric-comp,%
            natbib=true,backend=biber,maxbibnames=99,%
            giveninits=false,sorting=none]{biblatex}
\usepackage{nameref}
\usepackage{varioref}
\usepackage[noabbrev,capitalize]{cleveref}

\usepackage{etoc}
\usepackage{lineno}
\graphicspath{{figures/}}

\definecolor{doctodo}{HTML}{0D6EFD}
\definecolor{docnotelight}{HTML}{5A85D4}
\definecolor{docred}{HTML}{D62728}
\definecolor{dochl}{HTML}{FFF2CC}
\newif\ifshownotes
\shownotesfalse

\newcommand{\agentname}{visit preparation}
\newcommand{\Agentname}{Visit preparation}

\renewcommand{\abscontent}{}

\title{A prospective randomized study of patient-centered AI for clinical visit preparation}
\author[1]{Xincheng Tan}
\author[1]{Matthew Thompson}
\author[1]{Xiang Ji}
\author[1]{Angelica Willis}
\author[1]{Will Vaughan}
\author[1]{Roma Ruparel}
\author[1]{\\Michael A. Sanchez}
\author[1]{Sean Chang}
\author[1]{Joseph Xu}
\author[1]{Bhavna Daryani}
\author[1]{Preeti Singh}
\author[1]{Pinal Bavishi}
\author[1]{Kejia Chen}
\author[1]{\\Fan Zhang}
\author[1]{Priya Gupta}
\author[1]{Ernest Lim}
\author[1]{Christina Chen}
\author[1]{Bob Lou}
\author[1]{Dimitrios Antos}
\author[1]{Zach Wasson}
\author[1]{Ray Luo}
\author[1]{Fadi Yousif}
\author[1]{Florence Thng}
\author[1]{Mark Malhotra}
\author[1]{Joseph Breda}
\author[1]{Daniel McDuff}
\author[1]{Kavi Goel}
\author[1]{Katherine Chou}
\author[1]{Dale R. Webster}
\author[1]{Avinatan Hassidim}
\author[1]{Yossi Matias}
\author[1]{Po-Hsuan Cameron Chen}
\author[1]{Yun Liu}
\author[$\dagger$,1]{\\Mike Schaekermann}
\author[$\dagger$,1]{Quang Duong}
\author[$\dagger$,1]{Jackie M. Barr}

\affil[$\dagger$]{Equal leadership}
\affil[1]{Google Research}

\correspondingauthor{$\{$caratan$,\,$mikeshake$\}$@google.com}

\begin{document}

\begin{refsection}

% One-paragraph abstract.
% Character count: 1918 characters (of 1920 arXiv limit)

\begin{abstract}
Patients often arrive at brief, time-constrained clinical visits unsure of what to expect or ask, and millions of adults now turn to general-purpose AI chatbots to prepare beforehand. Yet whether conversational AI improves visit readiness remains unclear. We designed a large language model (LLM)-based visit preparation agent to guide patients before an appointment by gathering clinical context, setting visit expectations, providing relevant health information, building a prioritized list of questions for their clinician, and generating a summary. In a prospective consented randomized study in the Google Health app, 802 U.S. adults held 866 health-related conversations with the visit preparation agent (standard or concise variant) or a minimally prompted control chatbot built on the same LLM. The visit preparation agent significantly outperformed the control across blinded clinician evaluations and patient self-reports. In clinician reviews of completed sessions ($n = 272$), the agent achieved a 16.6\% higher rate of comprehensive preparation (67.2\% vs. 50.6\%; $P < 0.05$) and an 8.5\% higher rate of alignment with patients' health context (98.9\% vs. 90.4\%; $P < 0.01$), while maintaining high medical safety and appropriateness across all study arms. Patient post-surveys ($n = 460$) showed 16.8\% higher visit preparedness (73.8\% vs. 57.0\%; $P < 0.01$), 14.3\% higher confidence in knowing what to ask (81.4\% vs. 67.1\%; $P < 0.01$), and 12.0\% higher overall satisfaction (74.0\% vs. 62.0\%; $P < 0.05$). Clinician and participant metrics were not significantly correlated, indicating that expert review and patient self-assessment capture complementary dimensions of patient-facing AI. These findings suggest that LLM chatbots can safely help patients prepare for clinical visits, with greater benefit when following a structured workflow, and motivate future studies on in-visit communication and clinical outcomes.
\end{abstract}

\maketitle
{\noindent\absfont\theabstract\par}

% ------------ MAIN SECTIONS ---------------------
\section{Introduction}
\label{sec:introduction}

\begin{figure}[p]
  \centering
  \includegraphics[width=0.86\textwidth]{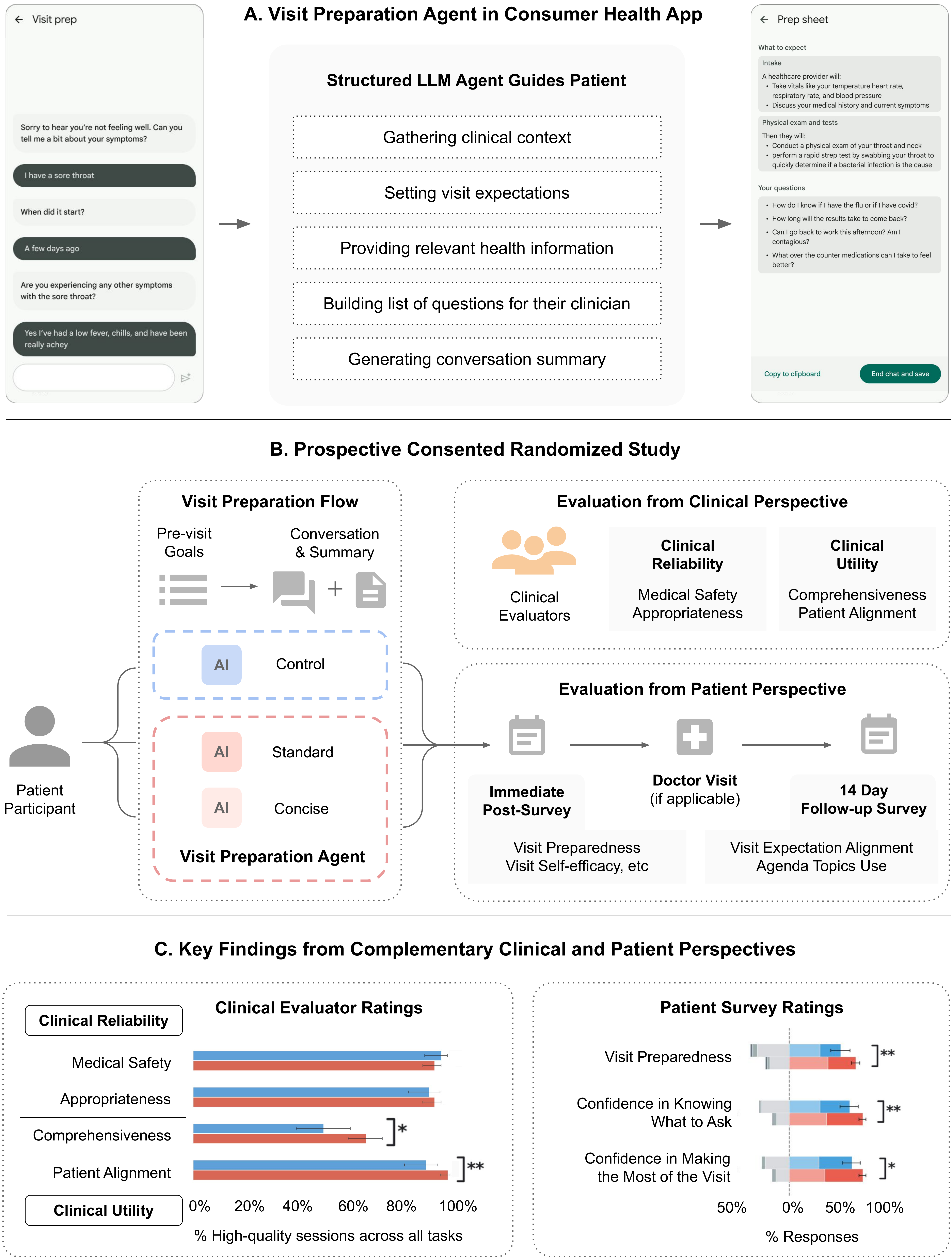}
  \vspace{0.1cm}
  \caption{\textbf{Overview of Contributions.} This work provides the first prospective assessment of AI-based visit preparation in a consumer health application, enabled by three key contributions: \textbf{(A)} A structured visit preparation agent powered by Gemini designed to guide patients through a structured pre-visit preparation flow (the user interface shown is trimmed for brevity and clarity; see \cref{fig:user_interface} for more details); \textbf{(B)} A prospective randomized study comparing the visit preparation agent with a control LLM agent; \textbf{(C)} Clinical evaluation suggesting increased clinical utility while maintaining high clinical reliability, with increases in patient-reported visit preparedness and visit self-efficacy.}
  \label{fig:overview}
\end{figure}

A growing number of patients are turning to artificial intelligence (AI) to navigate their healthcare, with recent national polling indicating that 1 in 4 U.S.\ adults use AI tools for health advice, among whom about 6 in 10 explicitly leverage AI to research conditions and formulate thoughts prior to a medical consultation~\cite{saad2026americans}.
This growing use of pre-consultation digital tools is happening in the backdrop of the realities of modern healthcare, where widespread clinician shortages~\cite{aamc2024complexities} and rising administrative demands~\cite{sinsky2016allocation,shanafelt2022changes} create time pressure on face-to-face clinical encounters.
In the United States, routine primary care visits are typically limited to 15 to 20 minutes~\cite{taiseale2007time}, and in health systems such as the United Kingdom and Japan, appointments are often constrained to 10 minutes or even less~\cite{irving2017international,kato2019building,dedumast2024trends}.
Within these brief encounters, patients with limited health literacy~\cite{kutner2006health,berkman2011low} and visit-related anxiety often arrive unsure of what to expect and struggle to articulate their concerns, recall critical details, or prioritize questions on the spot.

Long before the emergence of patient-facing AI tools, clinical communication research established that effective visit preparation is a catalyst for a productive doctor visit, heavily correlated with improved diagnostic accuracy, treatment adherence, and collaborative synergy between the patient and the clinician~\cite{street2009communication,zolnierek2009physician}.
Clinicians have long observed and advocated that structured pre-visit preparation by patients directly enhances in-visit communication efficiency, shared decision-making, and mutual satisfaction~\cite{elwyn2012shared}.
However, traditional pre-visit preparation paradigms have primarily relied on static behavioral interventions such as Question Prompt Lists (QPLs)~\cite{brandes2015characteristics,kinnersley2008interventions} and self-guided digital search tools.
Despite their documented utility in increasing patient question-asking and active engagement~\cite{kinnersley2008interventions,dimoska2008prompt}, standard QPLs require significant clinical time and resources to construct and are inherently static, while self-directed web search to research topics in advance shifts the burden of interpreting complex information and planning the visit entirely onto the patient.
Both approaches often presuppose adequate health literacy or a confirmed diagnosis, which may not apply in many clinical situations~\cite{ramlakhan2023optimizing,tracy2020exploring}.

Against these limitations, medical educators and practicing clinicians note in recent commentary that patients are now proactively turning to AI-assisted pre-visit reflection to make the most of their limited appointment time and regain agency within a strained system~\cite{rodman2026take}.
Recent advances in large language models (LLMs) offer promising foundational capabilities to support this emerging patient behavior in a conversational and intuitive manner.
In both static benchmarks and simulated clinical settings, LLMs have demonstrated strong medical reasoning across complex clinical tasks, such as generating differential diagnoses and drafting management plans that reach or surpass clinician-level performance~\cite{singhal2023large,nori2023capabilities,eriksen2023use,tu2024conversational}.
However, despite millions of patients already using general-purpose AI before their appointments~\cite{saad2026americans}, the safety, helpfulness, and usability of LLMs conversing directly with patients to offer structured pre-visit preparation guidance has not been rigorously studied.

In this study, we developed and conducted a randomized study using a visit preparation agent built to leverage a structured flow (including context gathering, expectation setting, health information provision, agenda topic suggestion, and summarization) and evaluated it from two complementary perspectives.
The primary outcome was the patient-reported preparation experience, encompassing their visit preparedness, self-efficacy, satisfaction, and likelihood of reuse, with an exploratory 14-day follow-up survey. Notably, because recruited participants self-reported wanting to prepare for a doctor's visit, these sentiments likely reflect real-world reactions of a group of patients prior to an encounter.
The secondary outcome was clinician-assessed quality of agent-generated content, evaluated by independent clinicians along axes assessing medical safety, appropriateness, comprehensiveness, and patient alignment.

Our key contributions are (\cref{fig:overview}):
\begin{itemize}
    \item \textbf{A randomized online study at scale:} a three-arm randomized study offered to 5,960 adults living in the United States on the Google Health App that compares a minimally-prompted control chatbot with two variants of a structured \agentname{} agent.
    \item \textbf{Dual-perspective evaluation:} independent clinicians assessed the medical safety, appropriateness, comprehensiveness, and patient alignment of each predefined preparation step; participants reported their visit preparedness, self-efficacy, satisfaction, and likelihood of reuse in a survey immediately after the conversation.
    \item \textbf{Key findings:} in the clinical review, conversations with the \agentname{} agents were rated more comprehensive and better aligned with each patient's health context, while medical safety and appropriateness was similarly high compared with Control. Participants who used the \agentname{} agents also more often reported feeling prepared for their visit, confident about what to ask, feeling satisfied with the tool, and wanting to use the tool again for a future visit.
\end{itemize}
\section{Methods}
\label{sec:methods}

\subsection{Study Design and Oversight}
\label{sec:methods:study_design}

\begin{figure}[ht!]
    \centering
    \includegraphics[width=0.9\textwidth]{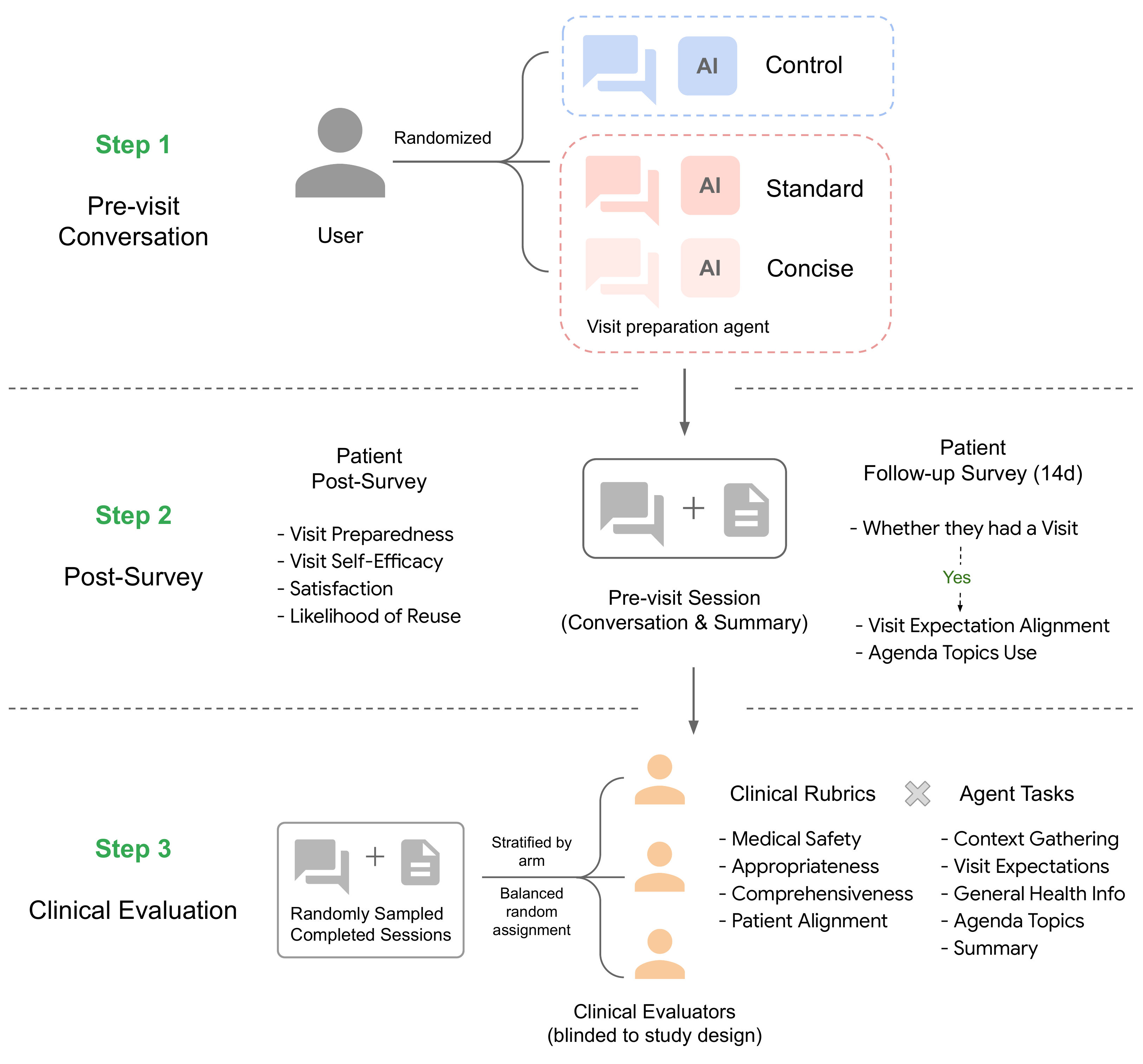}
    \vspace{0.1cm}
    \caption{\textbf{Study Design.} \textbf{Step 1:} Participants were randomized 1:2:2 to a minimally prompted base LLM (Control) or to one of two structured \agentname{} agents (standard or concise) for a pre-visit conversation. All three agents used the same LLM and interface, differed only in their system prompts, and each maintained a pre-visit summary (full prompts in \cref{appendix:agent_prompts}). \textbf{Step 2:} After the session, participants rated their visit preparedness, visit self-efficacy, satisfaction, and likeliness of reuse in a post-survey, and 14 days later were invited to a follow-up survey on whether they had a visit, whether the visit matched the expectations discussed, and whether they used the prep topics. \textbf{Step 3:} In a separate retrospective evaluation, a random sample of completed sessions (conversation and summary), stratified by arm, was assigned at random in a balanced manner to three PCP clinicians blinded to the study design and LLM associated with the conversation, who rated each agent task on four clinical rubrics.}
    \label{fig:study_design}
\end{figure}

We conducted a prospective, three-arm, randomized study to test whether a structured, large language model (LLM)-powered visit preparation agent helps patients prepare for an upcoming clinical visit better than a minimally prompted LLM chatbot. Results from the two structured LLM visit preparation agents are reported together due to demonstrating similar results (see below for more details), with dis-aggregated results presented in the appendix (\cref{appendix:arm_comparison}).
The study was conducted decentralized (online) via the Google Health mobile application in the United States from November 19, 2025, to May 15, 2026 (\cref{fig:study_design}).
This study was approved by the Advarra Institutional Review Board (IRB) (Protocol No. Pro00089880), and all participants provided electronic informed consent before enrollment.
The study was registered at ClinicalTrials.gov (NCT07552311) and reporting followed the Consolidated Standards of Reporting Trials (CONSORT) guidelines.

\subsection{Participants}
\label{sec:methods:participants}

Eligible participants were English-speaking adults aged 18 years or older who lived in the United States, and used the Google Health App on an Android smartphone.
We excluded participants with known mental health disorders, sexually transmitted diseases or infections, or pregnancy at the time of enrollment.

Analyses followed a modified intention-to-treat (mITT) approach, in which the primary analysis cohort comprised participants who had at least one substantive, health-related conversation with their assigned agent after randomization.
We excluded participants who never interacted with the agent, along with conversations that contained no health-related topics (\cref{appendix:prompt_noisy_convo}) or were not initiated directly through the visit-preparation entry point.
We also excluded conversations from participants who did not meet the eligibility criteria above despite self certification. This could have been due to the time period between consent (which enables use of the LLM agents) and the actual usage of the agents for pre-visit preparation; post-enrollment interactions that discussed symptoms of the excluded condition categories were further filtered via LLM-based flagging (\cref{appendix:prompt_eligibility_check}) followed by manual clinician review of each flagged case.

\subsection{Randomization and Interventions}
\label{sec:methods:interventions}

Participants were randomized 1:2:2 to one of the three agents: a minimally prompted control agent (Control), a structured \agentname{} agent with standard-length responses (standard), or a concise variant of the \agentname{} agent (concise).
The unequal allocation was intended to increase numbers of AI conversations available for analysis.
All three agents used the same underlying LLM (Gemini 2.5 Flash) and the same study interface in the Google Health App; they differed only in their system prompts (not visible to the participant), summarized as follows (instruction-level comparison in \cref{fig:supp_instruction_comparison}; full prompts in \cref{appendix:agent_prompts}).

\textbf{Control.}
The control agent received a minimal system prompt to respond to the user's health questions and stay on health-related topics, and refrain from giving medical advice, diagnoses, or treatment recommendations.
It received no additional guidance on how to prepare the participant for a clinical visit, other than to maintain a structured pre-visit summary at each turn covering relevant sections of a typical visit preparation conversation. This minimal prompt was intended to respect study design boundaries while approximating what the average user would experience without  specifically prompting their chatbot.

\textbf{\Agentname{} agents.}
We designed the \agentname{} agents to follow a structured, multi-step visit preparation workflow, prioritized according to the preparation goals the participant selected at the start of the session, including:
(1) \textit{context gathering} (type of visit, health needs, symptoms, and relevant medical history, etc.), scoped to the information needed to tailor the later preparation steps rather than a full clinical history;
(2) \textit{visit expectation setting} (what is likely to happen, such as questions the clinician may ask or tests that may be ordered);
(3) \textit{general health information} (answering general health inquiry in plain language, and explaining medical concepts);
(4) \textit{agenda topic generation} (the questions and topics most worth raising, including alternatives, next steps, and ``what-if'' scenarios); and
(5) \textit{summarization} (the pre-visit summary, updated after every turn).

The main comparison in this paper contrasted Control with the two \agentname{} arms combined, as both implemented the same structured workflow.
In a secondary analysis, the three arms were compared pairwise, including the two \agentname{} variants with each other (\cref{appendix:arm_comparison}).

\subsection{Outcomes}
\label{sec:methods:outcomes}

The primary outcome was participants' assessment of their preparation experience, reported immediately after the conversation.
The secondary outcome was a retrospective review of the agent-generated content by independent clinicians, who rated its clinical quality against predefined rubrics.
Exploratory outcomes covered the participant's reported visit experience at day 14 after the initial preparation conversation.

\subsubsection{Clinical Evaluation}
\label{sec:methods:clinical_evaluation}

Clinical evaluation was conducted retrospectively via stratified random assignment among three board-certified primary care clinicians with at least 7 years of post-residency clinical experience. These independent clinicians were blinded to the study design, the LLM associated with the conversation, and any information beyond what was shared in the conversation. 

Sessions under evaluation were restricted to the ones in which participants completed the visit preparation flow, i.e., the discussion covered all the goals the participant specified in the beginning.
From these sessions, we randomly sampled up to 102 meaningful interactions per arm.
For each arm, the sampled interactions were randomly and evenly assigned to one of the three evaluators.
Sessions identified as ineligible during the eligibility check (\cref{sec:methods:participants}) were removed.

For each session, the clinician reviewed the full conversation and the final pre-visit summary side by side, and rated every applicable preparation step (context gathering, visit expectation setting, general health information, agenda topics, and summary) on four clinical dimensions: \textit{medical safety}, \textit{appropriateness}, \textit{comprehensiveness}, and \textit{patient alignment} (\cref{tab:clinical_rubrics}).
Every dimension other than patient alignment applies to every step, while patient alignment was assessed only for agenda topics and the summary.
Each rating used a 5-point Likert scale (1 = lowest quality; 5 = highest quality), and a rating of 4 or 5 was considered a clearly \textit{positive} sentiment.

\subsubsection{Patient-Reported Outcomes}
\label{sec:methods:patient_reported_outcomes}

Immediately after the conversation, participants were invited to complete a brief survey to report their visit preparedness and experience across five aspects, each on a 5-point Likert scale (1 = most negative; 5 = most positive):
\textit{visit preparedness} (how prepared the participant felt for the upcoming visit);
\textit{confidence in knowing what to ask} (confidence about which questions and topics to raise with the clinician);
\textit{confidence in making the most of the visit};
\textit{satisfaction} with the tool; and
\textit{likelihood of reuse} for future visits.
A rating of 4 or 5 was considered \textit{positive}.

Participants were also invited via an in-app notification, 14 days after the conversation, to report in a follow-up survey whether they had a doctor visit in the past 2 weeks. If they did, they could further rate how well the visit matched the expectations discussed with the agent (\textit{visit expectation alignment}, 5-point Likert scale) and whether they used the agenda topics during the visit (\textit{agenda topic use}).

\subsection{Statistical Analysis}
\label{sec:methods:statistical_analysis}

All outcomes were analyzed as the proportion of sessions with a positive rating, reported with 95\% Wilson confidence intervals.
Arms were compared using two-sided two-proportion $z$ tests and Mann-Whitney U test at a significance level of 0.05.
To account for multiple comparisons, we controlled the false discovery rate using the Benjamini--Hochberg (BH) procedure under $q \leq 0.05$ within prespecified families of tests, as described below.

The main comparisons between arms were performed across all eligible sessions. To account for potential non-independence due to repeated participation (multiple chat sessions by the same participant for a small subset), we conducted a secondary sensitivity analysis restricting the dataset to each participant's first eligible session, and reported our findings in the appendix (\cref{appendix:non_independence_analysis}).

\subsubsection{Clinical Evaluation}
\label{sec:methods:stats_clinical}
We formulated endpoints at two analytical levels: 4 dimension-level endpoints capturing aggregated performance across all preparation steps (the percentage of sessions in which every applicable step received a positive rating on a given dimension), and 17 step-level endpoints tracking performance within individual steps (the percentage of sessions receiving a positive rating on a given dimension for a specific step). Between-arm differences were evaluated using two-sided, two-proportion $z$-tests for dimension-level endpoints and Mann–Whitney $U$ tests (on the Likert-scale ratings) for step-level endpoints, with $p$-values adjusted via the Benjamini–Hochberg (BH) procedure to control the false discovery rate across and within clinical dimensions, respectively.

To examine whether gains in utility came at the expense of clinical reliability, each session was cross-classified along two binary composite indicators: \textit{clinical reliability} (positive ratings across all safety and appropriateness items) and \textit{clinical utility} (positive ratings across all comprehensiveness and patient alignment items). Sessions were thus categorized into four mutually exclusive groups based on joint reliability and utility status. Between-arm differences in the proportion of sessions within each group were evaluated using uncorrected two-sided, two-proportion $z$-tests.

\subsubsection{Patient-Reported Outcomes}
\label{sec:methods:stats_pro}
For each of the five immediate survey items, we compared the proportion of positive responses (rating $\geq 4$) between arms using two-sided, two-proportion $z$ tests adjusted with BH procedure across two families: visit readiness \& confidence (\textit{visit preparedness}, \textit{confidence in knowing what to ask}, \textit{confidence in making the most of the visit}), and user experience (\textit{satisfaction} and \textit{likeliness of reuse}).

Follow-up surveys on \textit{visit expectation alignment} and \textit{agenda topic use} only included respondents who had a doctor visit within the past two weeks. 
Both were tested with two-sided, two-proportion $z$ test but without BH correction since the two items capture two different outcomes.

\subsubsection{Factors Associated With Positive Patient-Reported Outcomes}
\label{sec:methods:stats_factors}

To identify characteristics associated with positive visit preparedness and user experience, we fitted multivariable logistic regression models separately for each immediate survey item, defining the binary outcome as a positive rating (score of 4 or 5).
Models included the assigned agent and prespecified covariates adjusting for participant demographics (age and sex) and conversation content (visit type, disease area, and the types of visit expectations and agenda topics discussed). Full variable definitions are provided in \cref{appendix:factor_definitions}.
We report adjusted odds ratios (aORs) with 95\% confidence intervals and Wald test $P$ values.
Given that this analysis was exploratory, $P$ values were not adjusted for multiple comparisons and should be interpreted as hypothesis generating instead of validating.

\subsubsection{Exploratory Correlation Analyses}
\label{sec:methods:stats_correlation}

To examine the concordance between clinician and patient perspectives, we calculated Spearman rank correlation coefficients ($\rho$) with 95\% CIs between each survey item and clinical evaluation item among sessions with both ratings. To determine clinically meaningful alignment, correlations were tested against a prespecified threshold for moderate association ($\vert{}\rho\vert{} > 0.30$). Multiplicity was controlled per survey item across all clinical evaluation items using the BH procedure (\cref{appendix:joint_analysis}).

\section{Results}
\label{sec:results}

% ---------------------------------------------------------------------------
%                              Data Description
% ---------------------------------------------------------------------------
\subsection{Participant Characteristics}
\label{sec:results:participants}

\begin{figure}[tbp]
    \centering
    \includegraphics[width=\textwidth,height=0.74\textheight,keepaspectratio]{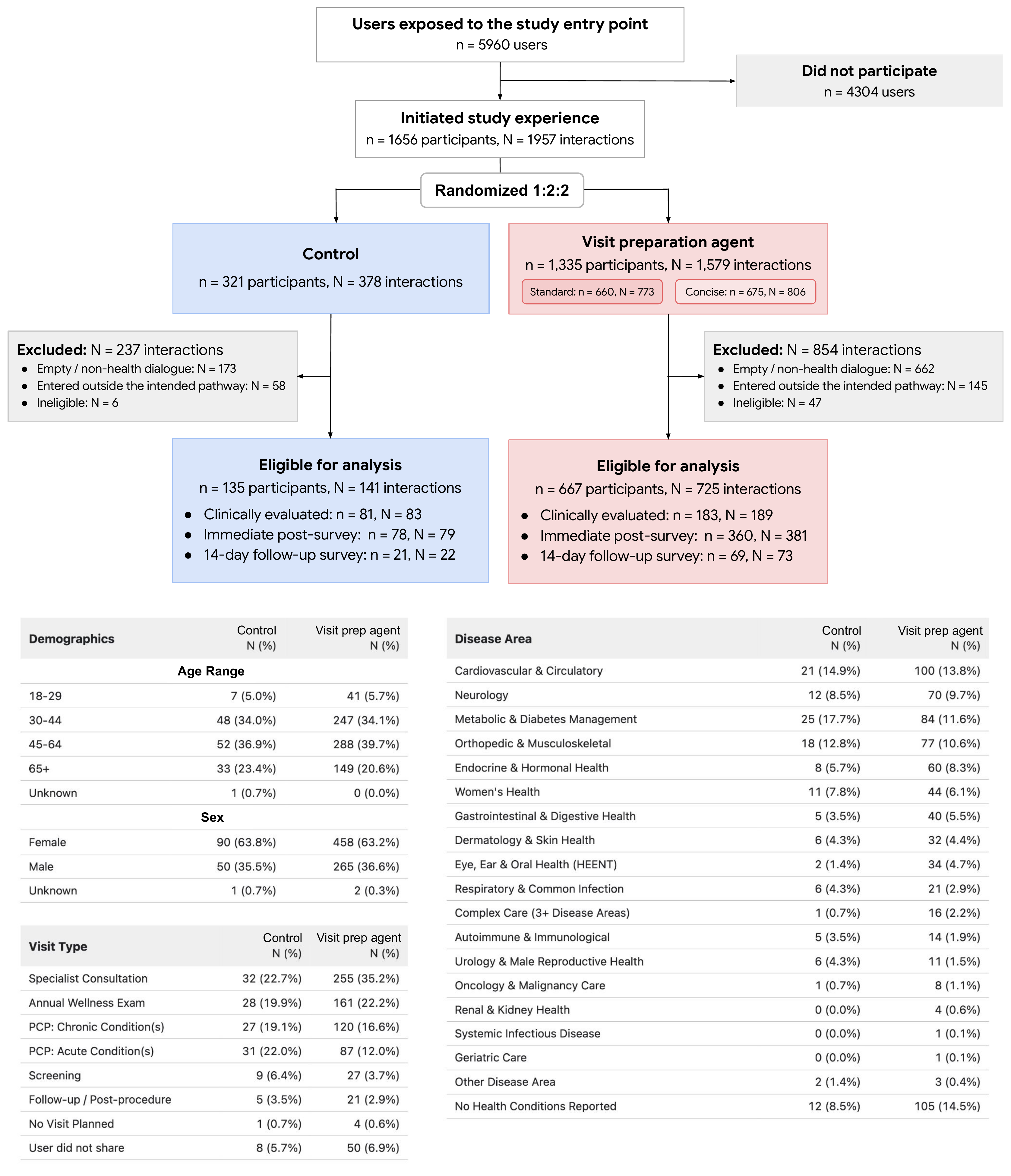}
    \vspace{0.1cm}
    \caption{\textbf{CONSORT Flow Diagram and Participant Characteristics by Arm.}
    Top: flow of participants (n) and conversations (N) from exposure to the study entry point, through 1:2:2 randomization to Control or to the standard or concise \agentname{} agent, to exclusions and inclusion in the analysis. Excluded conversations comprised empty or non-health dialogues, conversations entered outside the intended study pathway, and conversations centered on ineligible chief complaints identified by an LLM and verified via clinician review.
    In the primary analysis, the two \agentname{} intervention arms were pooled.
    Within each study arm, the analytic cohort reflects conversations that completed clinical evaluation, immediate post-survey, and 14-day follow-up reports.
    Bottom: demographics (age range and sex), visit type, and disease area of the included conversations, shown as N (\%) by arm. Visit type and disease area were parsed via LLM from each conversation.}
    \label{fig:participants}
\end{figure}

A total of 5,960 participants entered the in-app study entry point and were consented and randomized. Among these, 1,656 initiated the study experience, generating 1,957 interactions (Control: 321 participants, 378 conversations; \Agentname{} arms: 1,335 participants, 1,579 conversations; \cref{fig:participants}).
After excluding empty or non-health dialogues (Control: 173; \Agentname{}: 662), conversations entered outside the intended pathway (Control: 58; \Agentname{}: 145), and conversations from participants found to be ineligible (Control: 6; \Agentname{}: 47), 802 unique participants contributed 866 health-related conversations to the analysis (Control: 135 participants, 141 conversations; \Agentname{}: 667 participants, 725 conversations).
Across this cohort, participants completed the full preparation flow in 500 of the 866 conversations (57.7\%), with similar discontinuation rates between arms (Control: 41.1\%; \Agentname{}: 42.5\%). Among completed sessions, 272 conversations were sampled for blinded clinical evaluation (Control: 83; \Agentname{}: 189). 
For patient-reported outcomes, immediate post-survey responses were collected for 460 conversations (Control: 79; \Agentname{}: 381) and the 14-day follow-up survey reports for 95 conversations (Control: 22; \Agentname{}: 73).

Of the 866 included conversations, 548 (63.3\%) were held by female and 315 (36.4\%) by male participants, with sex unreported for 3 (0.3\%). Their mean age was 50.5 $\pm$ 14.5 years (Control: 51.0 $\pm$ 15.0; \Agentname{}: 50.4 $\pm$ 14.4), with age missing for 1 conversation.
Participants reported preparing for a wide range of visits, most commonly specialist consultations (287 [33.1\%]), annual wellness exams (189 [21.8\%]), primary care visits for chronic conditions (147 [17.0\%]), and primary care visits for acute concerns (118 [13.6\%]). The most frequent clinical domains were cardiovascular conditions (121 [14.0\%]), metabolic disorders and diabetes (109 [12.6\%]), musculoskeletal concerns (95 [11.0\%]), and neurology (82 [9.5\%]); 117 conversations (13.5\%) did not specify a health condition.
Participant demographics and disease domains distributions were broadly similar across arms (\cref{fig:participants}). However, visit types showed moderate differences: sessions in the \agentname{} arm more frequently addressed specialist consultations (35.2\% vs.\ 22.7\%) and less frequently addressed acute primary care concerns (12.0\% vs.\ 22.0\%) compared to control.

% ---------------------------------------------------------------------------
%                              Clinical Evaluation
% ---------------------------------------------------------------------------
\subsection{Clinical Evaluation}
\label{sec:results:clinical_evaluation}

\subsubsection{High clinical reliability and safety are maintained across study arms}
\label{sec:results:reliability}
\begin{figure}[ht!]
    \centering
    \includegraphics[width=\textwidth]{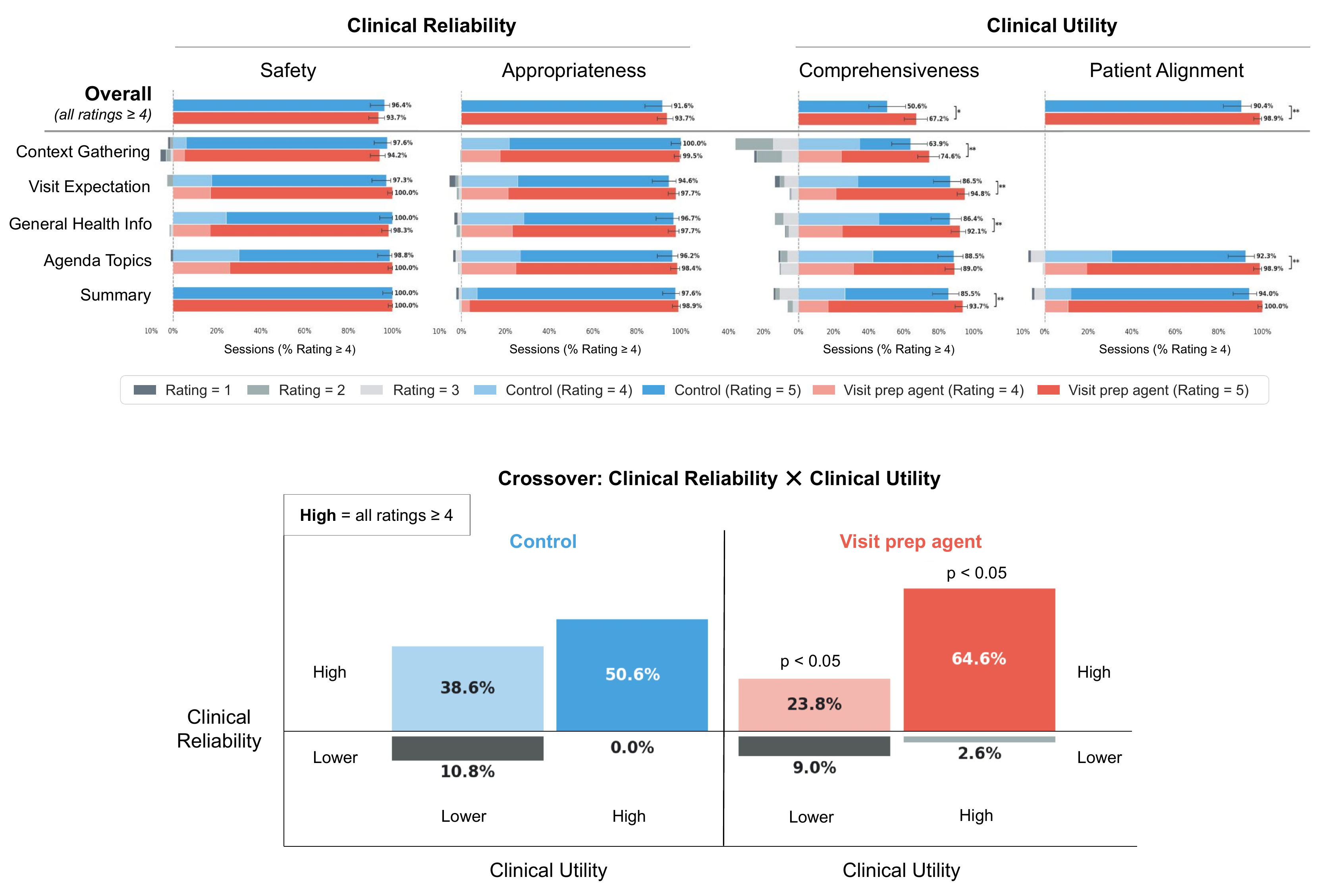}
    \vspace{0.1cm}
    \caption{\textbf{Clinical Evaluation Results.} \textbf{Top:} Share of sessions rated favorably (rating of 4 or 5) on each clinical dimension, overall (all applicable steps rated favorably) and for each preparation step. Bars to the left of 0\% show ratings of 1--3. The \agentname{} arm combines the concise and standard arms. Error bars indicate 95\% Wilson confidence intervals. Asterisks mark significant differences between the \agentname{} and Control arms ($^{*}P < 0.05$; $^{**}P < 0.01$): for the overall rows, in the share of sessions rated favorably (two-sided two-proportion $z$ test); for individual preparation steps, in how ratings were distributed across the 1--5 scale (Mann--Whitney U test). $P$ values are BH-adjusted within each clinical dimension. \textbf{Bottom:} Sessions grouped by clinical reliability (medical safety and appropriateness) and clinical utility (comprehensiveness and patient alignment), where \textit{high} indicates that all ratings in that category were 4 or higher. $P$ values compare the two arms on the share of sessions that were both highly reliable and useful, and on the share that were highly reliable but not highly useful (two-sided two-proportion $z$ tests, unadjusted).}
    \label{fig:clinical_eval}
\end{figure}

Based on blinded clinician evaluation, we found that high baseline clinical reliability was maintained across both arms, with no significant differences in medical safety or clinical appropriateness (\cref{fig:clinical_eval}, top). Aggregated along each clinical dimension, over 90\% of sessions in both arms received a positive rating across all applicable steps for medical safety (Control: 96.4\%; \Agentname{}: 93.7\%) and clinical appropriateness (Control: 91.6\%; \Agentname{}: 93.7\%). 

Across all five preparation steps, the agent performance were high (>90\%), with no statistically significant differences observed in safety or appropriateness between the arms. Minor numerical variations in rated safety during context gathering (94.2\% for the \agentname{} agents vs.\ 97.6\% for Control) and general health information (98.3\% vs.\ 100.0\%) were balanced by higher safety during visit expectation setting (100.0\% vs.\ 97.3\%; all $P > 0.05$ after BH adjustment).

\subsubsection{Structured preparation enhances clinical comprehensiveness and patient alignment}
\label{sec:results:utility}
Against this backdrop of similar safety and appropriateness (defined here as clinical reliability), the \agentname{} agents demonstrated higher comprehensiveness and patient alignment (defined here as clinical utility). Aggregated across all steps using a strict overall metric (all ratings positive), conversations and summaries generated by the \agentname{} agents were significantly more likely to be comprehensive (67.2\% vs.\ 50.6\%; $P < 0.05$) and closely aligned with the patient’s clinical context and goals (98.9\% vs.\ 90.4\%; $P < 0.01$).

Examining individual stages revealed that comprehensiveness gains were heterogeneous: the \agentname{} agents statistically significantly outperformed Control across 4 of the 5 steps, including context gathering (74.6\% vs.\ 63.9\%; $P < 0.01$), expectation setting (94.8\% vs.\ 86.5\%; $P < 0.01$), general health information provision (92.1\% vs.\ 86.4\%; $P < 0.01$), and chat summarization (93.7\% vs.\ 85.5\%; $P < 0.01$), while agenda comprehensiveness was not statistically higher (89.0\% vs.\ 88.5\%). For patient alignment, the advantage was concentrated in agenda topics (98.9\% vs.\ 92.3\%; $P < 0.01$) alongside a non-statistically significant increase in summary alignment (100.0\% vs.\ 94.0\%).

\subsubsection{Expanded clinical utility is achieved without compromising baseline safety}
\label{sec:results:tradeoffs}
A central concern in patient-facing AI is the potential tension between communication and utility versus safety, specifically whether providing more proactive, detailed guidance increases the risk of clinically inappropriate advice \cite{lee2023benefits, thirunavukarasu2023large}. To evaluate whether such a trade-off occurred, we characterized sessions along these axes: \textit{clinical reliability} and \textit{clinical utility} (\cref{fig:clinical_eval}, bottom).

The proportion of highly reliable sessions remained similar between study arms (Control: 89.2\%; \Agentname{}: 88.4\%). The divergence occurred in the utility achieved within those safe sessions: while 38.6\% of Control interactions were reliable yet fell short on utility, this share declined to 23.8\% with the \agentname{} agents ($P < 0.05$), and sessions achieving both high reliability and high utility rose from 50.6\% to 64.6\% ($P < 0.05$). Importantly, sessions exhibiting high utility alongside compromised reliability (the hallmark of ``unsafe helpfulness'') was rare across both arms (Control: 0.0\%; \Agentname{}: 2.6\%). In aggregate, structured visit preparation systematically increased the number of helpful interactions without compromising safety.

% ---------------------------------------------------------------------------
%                              Patient Outcomes
% ---------------------------------------------------------------------------
\subsection{Patient-Reported Outcomes}
\label{sec:results:user_perceived_helpfulness}

\subsubsection{Patients felt more prepared and more confident about their visit}
\label{sec:results:survey}

\begin{figure}[ht!]
    \centering
    \includegraphics[width=\textwidth]{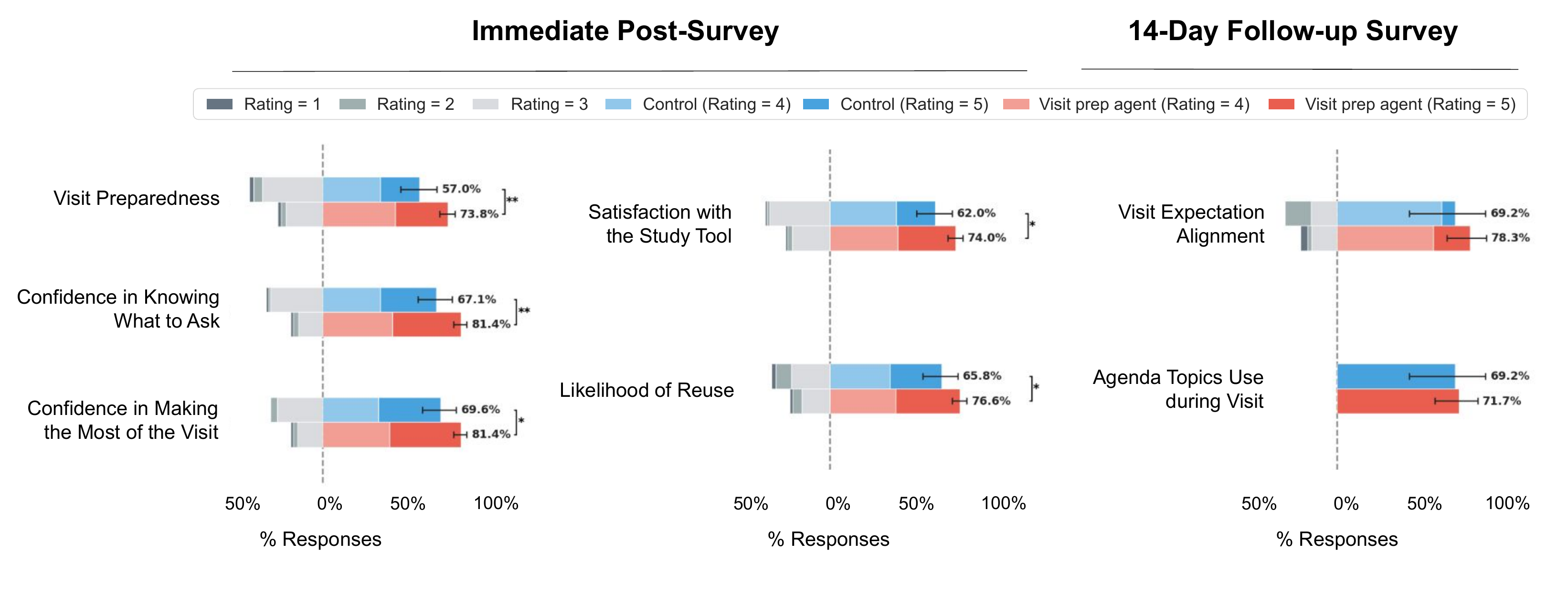}
    \vspace{0.1cm}
    \caption{\textbf{Patient-reported Outcomes.} Distribution of ratings in the immediate post-survey (left and middle) and the exploratory 14-day follow-up survey (right; limited to respondents who had a doctor visit: Control, 13; \Agentname{}, 46); the \agentname{} arm combines the concise and standard arms. Bars to the right of 0\% show positive responses (rating of 4 or 5; for agenda topic use, the share of participants who reported using them), with the total share of positive responses labeled; bars to the left show neutral and negative ratings. Error bars indicate 95\% Wilson CI. $^{*}P < 0.05$; $^{**}P < 0.01$ (two-proportion $z$ tests with BH correction across the five immediate survey items).}
    \label{fig:user_feedback}
\end{figure}

Participants who used the \agentname{} agents reported a better preparation experience than those who used the Control chatbot (\cref{fig:user_feedback}).
In the immediate post-survey, 73.8\% of the \agentname{} agent users reported feeling prepared for their upcoming visit, compared with 57.0\% of Control users ($P < 0.01$), a 16.8-percentage-point increase.
These participants were also more confident in knowing what to ask their clinician (81.4\% vs.\ 67.1\%; $P < 0.01$) and in making the most of the visit (81.4\% vs.\ 69.6\%; $P < 0.05$).
On the user experience end, they were also more satisfied with the tool (74.0\% vs.\ 62.0\%; $P < 0.05$) and more likely to want to use it again (76.6\% vs.\ 65.8\%; $P < 0.05$).

We received 95 responses to the 14-day follow-up survey ($n=22$ Control; $n=73$ \Agentname{}), of which 59 indicated that the participant had seen a clinician in the interim ($n=13$ Control; $n=46$ \Agentname{}). Among those who had a visit, 78.3\% of the \agentname{} group and 69.2\% of the Control group felt their appointment aligned with the expectations set beforehand by the agent, while 71.7\% and 69.2\%, respectively, reported raising their prepared agenda topics during the visit.
Neither difference reached statistical significance with the limited sample size of this follow-up cohort.

\subsubsection{Factors associated with visit preparedness and self-efficacy}
\label{sec:results:factors}

\begin{figure}[ht!]
    \centering
    \includegraphics[width=\textwidth]{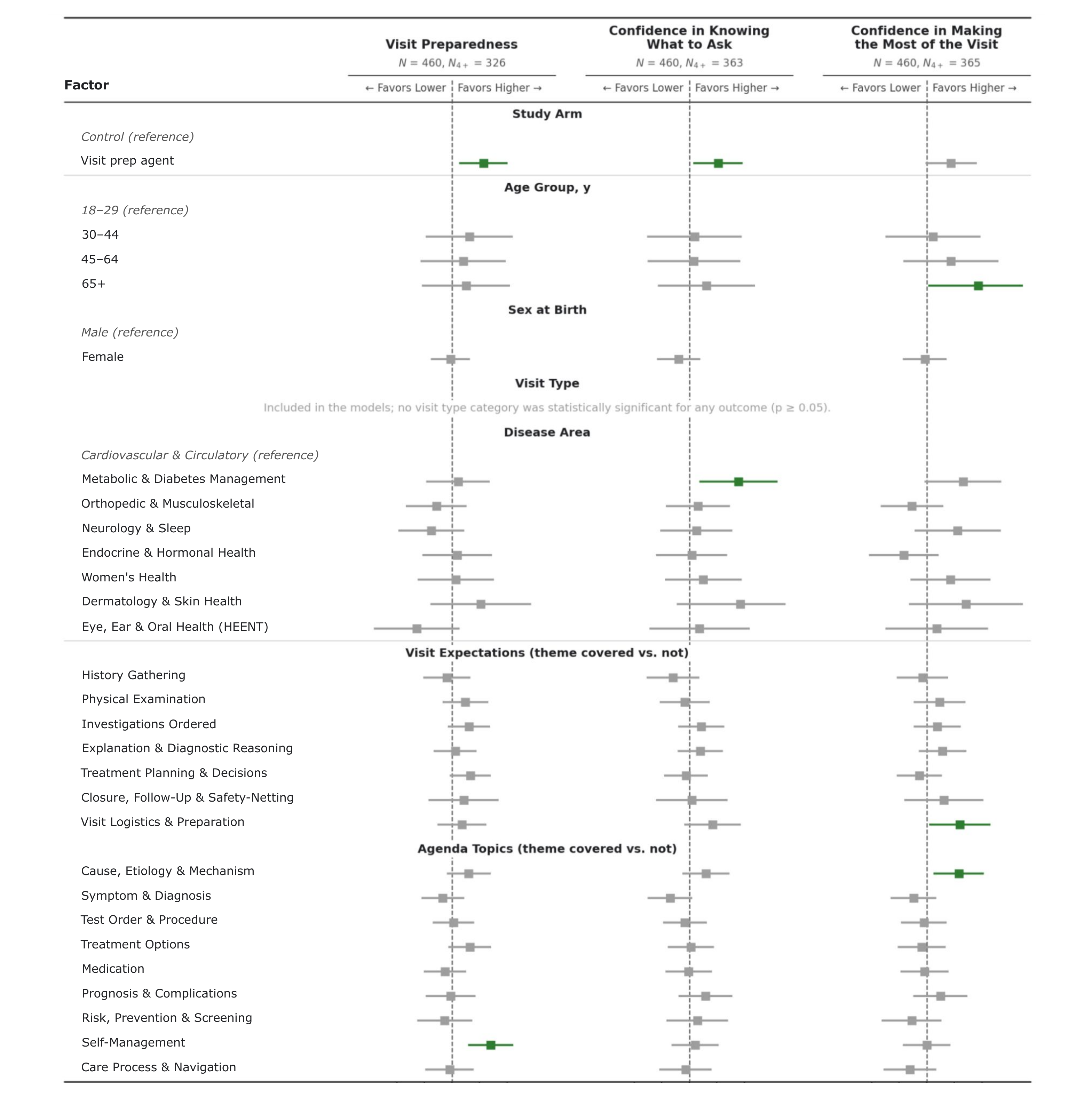}
    \vspace{0.1cm}
    \caption{\textbf{Factors Associated with Positive Visit Readiness and Confidence.} Adjusted odds ratios (log scale) with 95\% CIs from multivariable logistic regression of a positive response (rating of 4 or 5) for visit preparedness, confidence in knowing what to ask, and confidence in making the most of the visit. Each row is compared with the reference category in italics, or, for visit expectation and agenda topic themes, with conversations that did not cover the theme. Visit type was included in all models but is collapsed into a single row because no category reached significance for these outcomes. Green markers indicate $P < 0.05$ (Wald test, unadjusted for multiple comparisons). $N$, number of respondents; $N_{4+}$, number of positive responses. Full estimates, including satisfaction and likeliness of reuse, are reported in \cref{fig:supp_factor_analysis}.}
    \label{fig:factor_analysis}
\end{figure}

To explore what shapes a positive preparation experience, we fitted a multivariable logistic regression for each immediate survey item, adjusting for study arm, participant characteristics, visit context, and the content covered in the conversation (\cref{sec:methods:stats_factors}).
\Cref{fig:factor_analysis} presents the findings for the three primary visit-readiness and self-efficacy items; complete models, including analysis on satisfaction and likelihood of reuse, are detailed in \cref{appendix:factor_analysis}.

The benefit of the \agentname{} agents largely held after adjustment.
Participants who used the \agentname{} agents were significantly more likely than Control participants to report feeling prepared and knowing what to ask, and being satisfied with the tool.
The adjusted associations with confidence in making the most of the visit and with intent to reuse remained directionally consistent but did not reach statistical significance in this analysis.

Participant and visit context mattered only for specific outcomes.
Age 65 or older was significantly associated with having confidence to make the most of their visit, and likewise managing metabolic conditions or diabetes was associated with greater confidence in knowing what to ask.
Sex was not significantly associated with any survey outcome, and visit type was associated only with intent to reuse, which was higher among participants preparing for screening visits.

Conversational content was closely linked to participants' perceived support. Setting expectations around visit logistics and integrating disease etiology into the consultation agenda were both associated with greater confidence in making the most of the visit. Meanwhile, including self-management options on the agenda was correlated with higher feelings of visit preparedness. Taken together, preparation appeared most valuable when it helped participants understand possible diseases and how they work, and anticipate clinical decisions and concrete next steps. Given the exploratory nature of these analyses and the lack of adjustment for multiple comparisons, these associations should be interpreted as hypothesis-generating.

\subsection{Clinician and Patient Perspectives Were Complementary}
\label{sec:results:dual_perspective}

Finally, we tested the concordance between patients' impressions and clinicians' judgments of the same conversations (\cref{appendix:joint_analysis}).
After correcting for multiple comparisons, we found no significant correlation between any survey item and the clinical ratings of any preparation step, across all clinical dimensions (\cref{fig:supp_dual_perspective}).
Patient satisfaction, in other words, is not a proxy for clinical quality, and clinical quality does not guarantee that patients feel helped.
Evaluating patient-facing medical AI therefore requires both perspectives: independent clinical review to establish safety and quality, and patient-reported outcomes to establish value to the people the tool is meant to serve.
\section{Discussion}
\label{sec:discussion}

In this study, we developed and evaluated a LLM-based visit preparation agent deployed through a consumer mobile app.
The structured visit preparation agents demonstrated clinical reliability on par with a minimally instructed LLM chatbot while delivering significantly more comprehensive and personalized preparation guidance across multiple preparation steps.
Further analysis of the  clinically favorable sessions revealed that the agent performed particularly well for clinical safety and appropriateness, implying a promising baseline for future studies.
Among the axes assessed by clinicians, context gathering had the largest relative room for improvement. This is sensible from the clinical perspective; context gathering by the \agentname{} agents was not designed as a traditional clinical intake or history gathering function, which remains the clinician's role during the visit; rather, it aimed to collect just enough context to tailor visit expectations, health information, and agenda topics. That said, if one were to design a similar system to incorporate more directly into the downstream clinical workflow (such as handing over detailed clinical intake information to the clinician), the context gathering may need to be more extensive and lengthy.

From the patient perspective, a significantly higher proportion of participants in the \agentname{} arms reported feeling well-prepared for their upcoming visit including being prepared to ask their clinician questions and make the most of their visit after a structured visit preparation with the agent, compared to those in the control arm, accompanied by significantly higher satisfaction and intent to reuse for their future visit. Given that the control chatbot and the \agentname{} agents shared the same underlying model, these gains demonstrate that base model capability alone does not guarantee effective visit preparation; rather, how the interaction is structured is critical. Instead of leaving users to navigate open-ended responses on their own, the structured flow (context gathering, expectation setting, information provision, agenda topic suggestion, and summarization) actively guided patients to organize their concerns and build a concrete visit agenda prior to their appointment.

Crucially, however, after correction for multiple testing, user-perceived visit readiness and satisfaction showed no significant association with clinician-graded agent quality. This divergence echoes broader observations in patient-centered care research that subjective satisfaction often reflects the interpersonal experience rather than the technical quality of care received~\cite{street2009communication,rao2007communication}, implying that user-perceived helpfulness alone cannot replace rigorous evaluation of clinical soundness.
Evaluation frameworks that assess both objective clinical performance and subjective real user experience are therefore essential for more comprehensive, human-centered understanding of AI usefulness for health related purposes.

\subsection{Relation to Prior Work}
\label{sec:discussion:prior_work}

This study extends the nascent literature on LLM-based health applications in several ways.
Whereas prior work has primarily evaluated LLMs in clinical reasoning benchmarks~\cite{singhal2023large,nori2023capabilities,eriksen2023use} or in simulated diagnostic dialogue between AI and clinicians or standardized patients~\cite{tu2024conversational,hager2024evaluation}, our work focuses on the LLM in a fundamentally different role, as a patient-facing preparation guide prior to clinical encounters.
This pre-visit orientation shifts the objective from diagnosis to empowerment: helping patients organize \emph{their} thoughts, set what to expect at the visit, and arrive at their appointment better equipped for a productive interaction with their provider. Empowering patients and enhancing their clinical interactions is also not merely a one-time occurrence, given that patients interact multiple times per year over many years with health care providers. Our results suggest the potential for an ongoing supportive role for LLM within a patient's health care journey.

In some sense, the \agentname{} agents represent a modernized, interactive extension of the Question Prompt List (QPL) tradition~\cite{brandes2015characteristics,kinnersley2008interventions}.
QPLs have long been recognized as effective in increasing patient question-asking and engagement~\cite{sansoni2015question}, yet their routine use is limited by a resource-intensive development process, their static, condition-specific nature, and their reliance on a pre-existing diagnosis and adequate health literacy~\cite{clayton2003asking,ramlakhan2023optimizing,brandes2015characteristics,keinki2021effect,tracy2020exploring}.
The conversational agent approach addresses several of these barriers: it adapts dynamically to the user's specific health context, generates personalized agenda topics, and requires no clinician time to construct or distribute.
At the same time, the agent's structured flow (context gathering followed by staged information delivery) preserves the pedagogical scaffolding that makes QPLs effective while offering a more accessible and scalable delivery format, potentially for multiple health care visits over time..

The clinical safety profile observed in this study is also encouraging in the context of recent literature highlighting the limitations of LLMs in autonomous clinical decision-making~\cite{hager2024evaluation}.
Our clinical reliability-utility analysis showed that close to 90\% of sessions in both arms were rated highly reliable on every preparation step (Control: 89.2\%; \Agentname{}: 88.4\%), and that sessions combining high utility with lower reliability (the pattern that would signal unsafe helpfulness) were rare (\Agentname{}: 2.6\%).
These findings suggest that scoping LLMs into a structured, non-diagnostic visit preparation tool offers a low-risk pathway to empower patients and deliver meaningful clinical value.
% acceptable safety margins while still providing meaningful clinical value.

\subsection{Limitations}
\label{sec:discussion:limitations}

This study has several limitations.
First, not all included participants completed the full visit preparation flow (42.3\% of included conversations ended before completion), reducing the effective sample size and limiting statistical power for more granular subgroup analyses such as per disease area~\cite{eysenbach2005law}.
Because clinical evaluation was restricted to sessions that completed the flow, it may also be subject to survivorship bias: if some participants withdrew prematurely because of agent shortcomings (e.g., repetitive questioning or unaddressed concerns), clinical ratings may overestimate the quality of all conversations.
However, completion rates were similar between arms, which makes it less likely that this bias affected the between-arm comparison, but the absolute clinical ratings should be interpreted as reflecting completed conversations.
Another limitation is that our 14-day follow-up survey was intended to capture whether the effectiveness of visit preparation carried over to the actual encounter, but a relatively limited number of participants responded and even fewer had an encounter in that window.
Future studies could better evaluate the clinical impact by gathering feedback directly from participants' clinicians and understanding downstream health outcomes and other proxy measures of care quality.
Finally, our findings should be interpreted in the context of our specific study cohort, and may be influenced by unmeasured sociodemographic factors.
Because the participants were Android users who actively used the Google Health app and opted into a research study, they were likely more technologically engaged and health literate than the general patient population.
Future patient-centric AI studies should collect and account for detailed sociodemographic variables, including education level, socioeconomic status, primary language, race and ethnicity, and insurance status, to evaluate generalizability and ensure these tools benefit diverse populations.

\section{Conclusion}
\label{sec:conclusion}

This study demonstrates that a structured, LLM-based conversational agent can serve as a clinically safe and useful pre-visit preparation tool that meaningfully improves patient-reported visit preparedness and self-efficacy.
The dual-perspective evaluation confirms that specialized prompting enhances the breadth and personalization of preparation guidance without compromising medical safety, while also revealing that user-perceived helpfulness and clinical quality operate along independent dimensions, reinforcing the need for both perspectives in evaluation frameworks for patient-facing AI.
Future research should explore the end-to-end impact on visit experience and health outcomes from single and repeated real-world clinical encounters, incorporate provider sentiments, and investigate alternative interaction modalities to ultimately support longitudinal, personalized care companion experiences that extend beyond the pre-visit preparation step studied here.

\subsubsection*{Acknowledgments}
This project was an extensive collaboration between many teams at Google Research and Google Deepmind.
We thank Lin Yang and Yishay Mansour for their comprehensive review and detailed feedback on the manuscript.
We also thank Anil Palepu, Akshay Goel, Jimmy Li, Diego Ardila, Abbi Ward, Bhawesh Kumar, Jimmy Hu for their contributions in ideation of this work.
We are grateful to Catherine Kozlowski, Rachelle Sico, Jacinda Mein, Isha Mishra, Danielle Ziegelheim, and Jessica M. Williams for their review and support during our research.
We also thank Andreea Gorgoi, Adriana Draghici, Akekawit Jitprasert, Sam Spano, Jo Schaeffer, John Schmidt, Jasmin Valentin, Jennifer McHardy, Jasmine Weaver, Sheila Ramanathan, Sohrab Sidhu, and Heather Towery for their support in the study, as well as SiWai Man, Michelle Gatchalian, and Gordon Turner.
Finally, we are grateful for the support from Ines Mezerreg, Hieu Hoang, Jihyeon Lee, Alicia Mart\'{i}n, Qinghan Xue, Rory Sayres, Jeremy Lai, Xiang Yin, Jonathan Krause, and Per Bjornsson.

\subsubsection*{Data Availability}
Example user interface, conversation and session summary are provided in the manuscript and appendix (\cref{appendix:user_interface}).

\subsubsection*{Code Availability}
The study agents are research prototypes built on Gemini 2.5 Flash~\cite{comanici2025gemini} through system prompting; no model weights were trained or modified for this study.
For reproducibility, the system prompts of the three study arms are provided in \cref{appendix:agent_prompts}.

\subsubsection*{Competing Interests}
This study was funded by Alphabet Inc and/or a subsidiary thereof (`Alphabet'). Authors who are employees of Alphabet may own stock as part of the standard compensation package.

\newpage
\setlength\bibitemsep{3pt}
\printbibliography
\clearpage
\end{refsection}

% ------------ APPENDIX SECTIONS ---------------------
\begin{refsection}
\clearpage

\renewcommand{\thesection}{A.\arabic{section}}
\renewcommand{\thefigure}{A.\arabic{figure}}
\renewcommand{\thetable}{A.\arabic{table}}
\renewcommand{\theequation}{A.\arabic{equation}}
\renewcommand{\theHsection}{A\arabic{section}}
\renewcommand{\theHfigure}{A\arabic{figure}}
\renewcommand{\theHtable}{A\arabic{table}}

\setcounter{section}{0}
\setcounter{figure}{0}
\setcounter{table}{0}
\setcounter{equation}{0}

\noindent \textbf{\LARGE{Appendix}}\par
\normalfont

\section{User Interface and Example Conversation}
\label{appendix:user_interface}

\begin{figure}[h!]
    \centering
    \includegraphics[width=\textwidth]{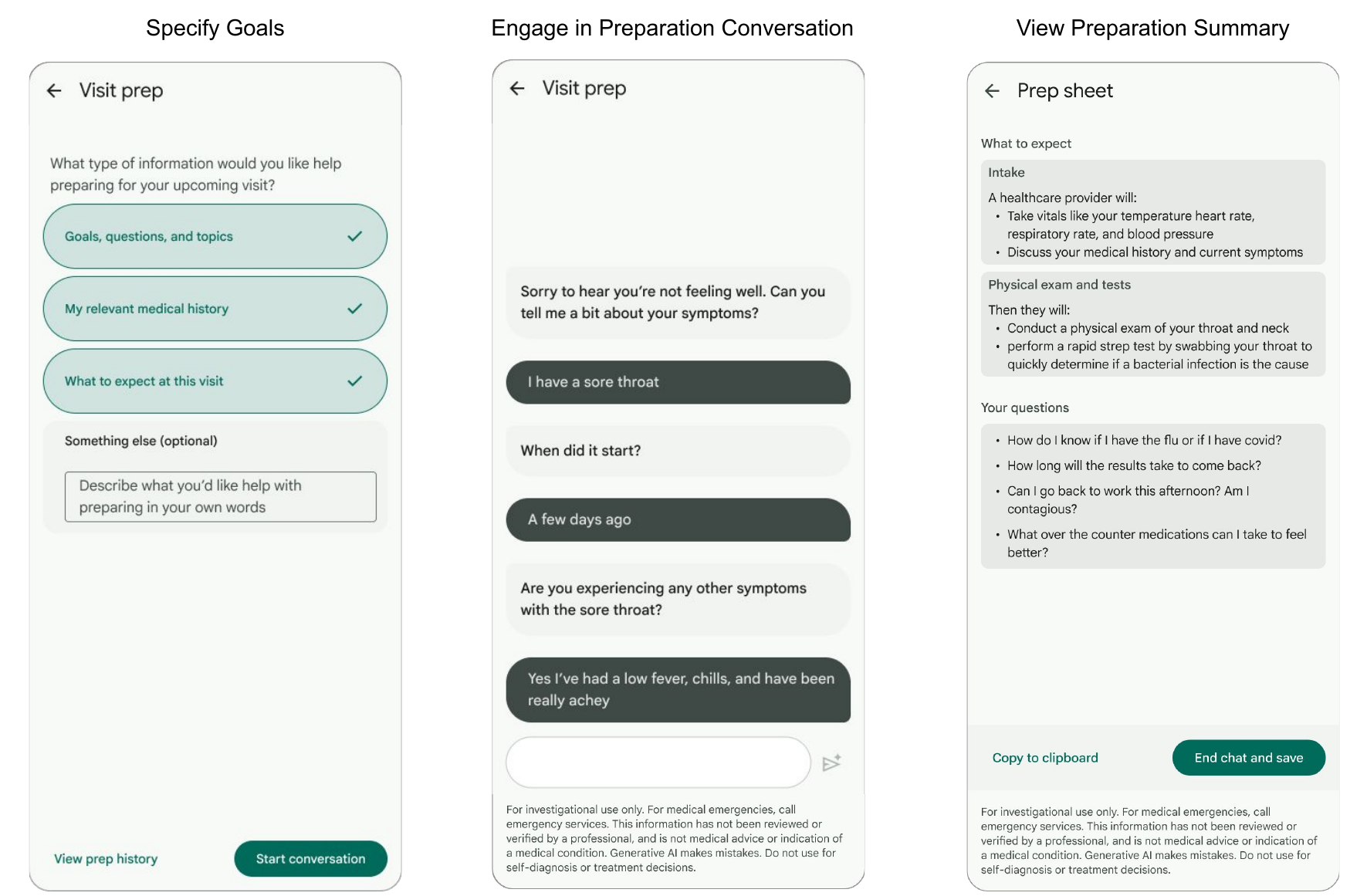}
    \vspace{0.1cm}
    \caption{\textbf{User Interface of the Study Tool in the Google Health App.} \textbf{Left:} Participants select their preparation goals and optionally describe what they would like help with in their own words. \textbf{Middle:} Participants converse with the agent in free text on their pre-visit preparation. \textbf{Right:} The pre-visit summary (prep sheet) organizes the key points of the conversation into a summary checklist, and can be copied or saved. The same interface was used in all three arms. Note: the conversation and summary shown are an illustrative example based on a synthetic scenario.}
    \label{fig:user_interface}
\end{figure}

All three study arms used the same interface in the Google Health App (\cref{fig:user_interface}); the arms differed only in the system prompt of the agent (\cref{sec:methods:interventions,appendix:agent_prompts}).
A session proceeded in three stages.
First, participants specified what they wanted help with in preparing for their upcoming visit by selecting one or more preparation goals (\textit{goals, questions, and topics}; \textit{my relevant medical history}; \textit{what to expect at this visit}) and, optionally, describing their needs in their own words.
Second, they engaged in a free-text conversation with the agent.
Third, they could view the pre-visit summary (\textit{prep sheet}), which was updated throughout the conversation and organized the information into sections such as what to expect at the visit and the questions to ask; participants could copy the summary to the clipboard or end the chat and save it for later reference.
The conversation and summary screens displayed a disclaimer that the tool was for investigational use only, that its output had not been reviewed by a professional and was not medical advice, and that participants should call emergency services in a medical emergency.

% \begin{mdframed}[backgroundcolor=black!5!white, linecolor=black!60!white, linewidth=1pt, roundcorner=1mm]
% \footnotesize
% \textbf{\large Example Conversation}

% \noindent\rule{\textwidth}{0.5pt}

% \noindent\textbf{\normalsize 1. Intent seeking}

% \noindent\textbf{User:} \docnote{<user turn>}\\
% \textbf{Agent:} \docnote{<agent turn>}

% \vspace{0.4em}
% \noindent\textbf{\normalsize 2. Context gathering}

% \noindent\textbf{User:} \docnote{<user turn>}\\
% \textbf{Agent:} \docnote{<agent turn>}

% \vspace{0.4em}
% \noindent\textbf{\normalsize 3. Visit expectations and general health information}

% \noindent\textbf{User:} \docnote{<user turn>}\\
% \textbf{Agent:} \docnote{<agent turn>}

% \vspace{0.4em}
% \noindent\textbf{\normalsize 4. Agenda topic discussion}

% \noindent\textbf{User:} \docnote{<user turn>}\\
% \textbf{Agent:} \docnote{<agent turn>}
% \end{mdframed}

% \vspace{0.4em}
% \begin{mdframed}[backgroundcolor=black!5!white, linecolor=black!60!white, linewidth=1pt, roundcorner=1mm]
% \footnotesize
% \textbf{\large Example Pre-Visit Summary}

% \noindent\rule{\textwidth}{0.5pt}

% \begin{itemize}[leftmargin=*, noitemsep, topsep=0pt, parsep=0pt, partopsep=0pt]
%     \item \textbf{Visit goals:} \docnote{<...>}
%     \item \textbf{Health context:} \docnote{<...>}
%     \item \textbf{What to expect at the visit:} \docnote{<...>}
%     \item \textbf{Agenda topics to discuss with your provider:} \docnote{<...>}
% \end{itemize}
% \end{mdframed}

\clearpage
\section{Agent System Prompts}
\label{appendix:agent_prompts}

This section reproduces the system prompts of the three study arms (\cref{sec:methods:interventions}) verbatim. All arms used the same underlying LLM and the same conversational interface, and differed only in their system prompts. Each arm, including Control, returned a structured pre-visit summary (JSON) at every agent turn alongside its conversational reply, which participants could view as a ``prep sheet''.

\begin{figure}[ht!]
    \centering
    \includegraphics[width=0.6\textwidth]{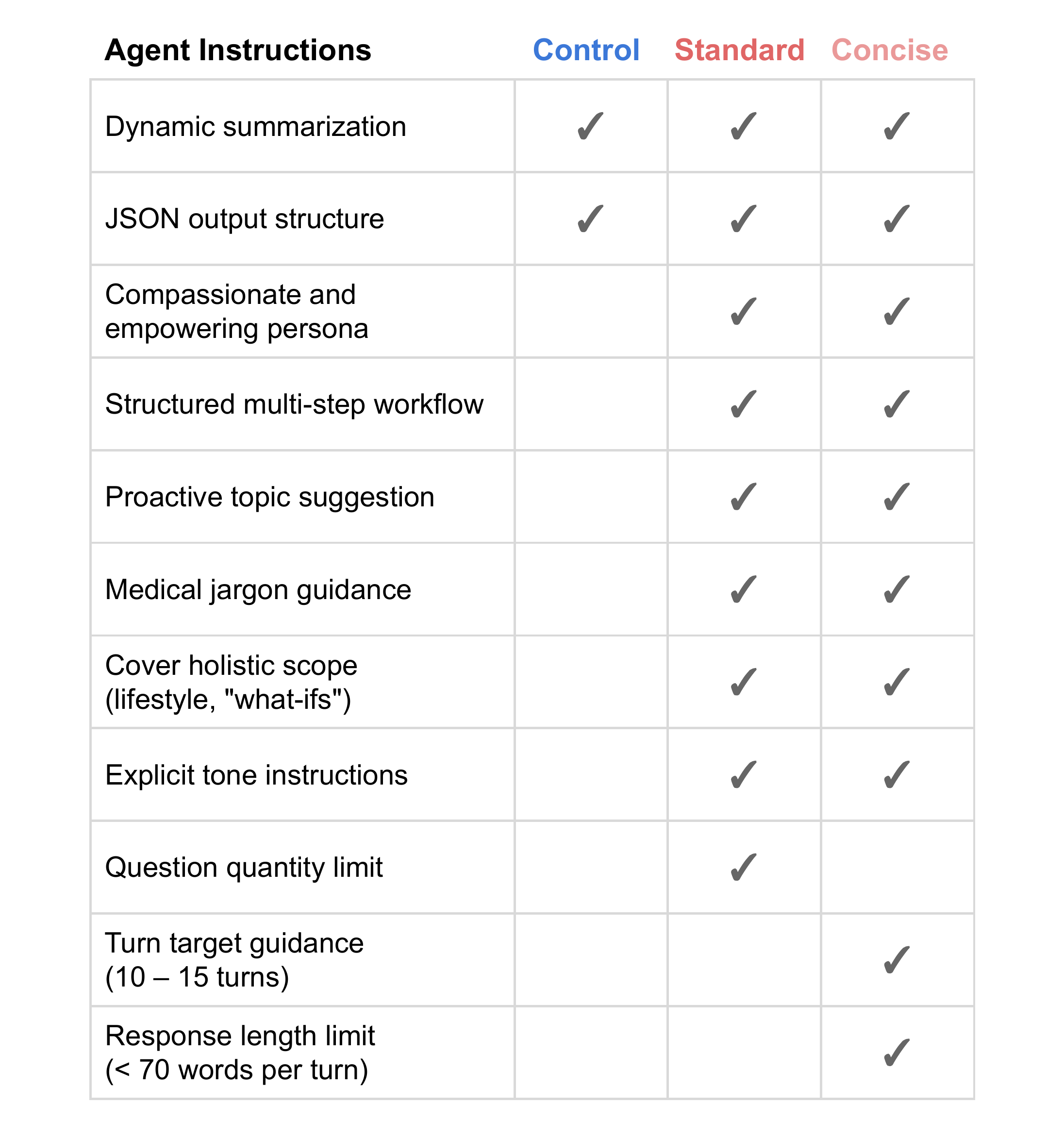}
    \vspace{0.1cm}
    \caption{\textbf{Comparison of Core Instructions Across Study Arms.} A check mark indicates that the instruction is included in the system prompt of the Control, standard \agentname{}, or concise \agentname{} agent. All three arms used the same underlying LLM and conversational interface; the full prompts are reproduced in \cref{appendix:prompt_control,appendix:prompt_standard,appendix:prompt_concise}.}
    \label{fig:supp_instruction_comparison}
\end{figure}

\Cref{fig:supp_instruction_comparison} compares the core instructions of the three arms, and the key differences are:
\begin{itemize}[leftmargin=*, noitemsep, topsep=2pt]
    \item \textbf{Control} (\cref{appendix:prompt_control}): a minimal prompt that lets the model respond as it normally would, restricts the conversation to health-related topics, states that the agent cannot offer medical advice, and specifies the pre-visit summary format. It prescribes no conversation flow.
    \item \textbf{Standard \agentname{} agent} (\cref{appendix:prompt_standard}): a structured 10-step visit preparation flow, prioritized by the goals the user selected before the conversation. The agent keeps each turn short and asks one question at a time, with no explicit word limit.
    \item \textbf{Concise \agentname{} agent} (\cref{appendix:prompt_concise}): the same 10-step flow, but with tighter pacing. Responses are ideally under 70 words, the conversation aims for 10--15 turns, 2--3 related questions may be grouped in a single turn, and the flow is used as a checklist rather than followed step by step. It also adds worked examples for discussing practical barriers such as cost, transport, or time (step~8) and for distinguishing expected symptoms from red-flag symptoms (step~9).
\end{itemize}

\subsection{Control}
\label{appendix:prompt_control}
\begin{lstlisting}[style=prompt]
The user will tell you symptoms or discuss their upcoming doctor visit. You can
respond as you normally would.

You should not converse on topics unrelated to answering health-related queries.
If the user attempts to converse on a topic unrelated to health-related queries,
you should kindly redirect the user back to asking about health-related queries.

You are *not* a medical professional and *cannot* offer medical advice,
diagnoses, or treatment recommendations.

## OUTPUT FORMAT

AT EACH TURN of the conversation, please summarize what has been discussed in
the following JSON format:

* `text_response` (String): Your main chat response to the user for the current
  turn.
*   `session_title` (String): A precise and descriptive title of the visit (no
    more than 5 words). Only start populating this once the user has provided
    some context of their visit, otherwise this field should be an empty string.
* `session_overview` (String): A one-sentence, **concise** summary of the
  conversation up to this point (ideally under 15 words). Only start populating
  this once the user has provided some context of their visit, otherwise this
  field should be an empty string.
* `visit_preparation_summary_json` (List[VisitPrepSummarySection]): A list of
  VisitPrepSummarySection objects, each covers key facts from the conversation
  so far, including your text response in the current turn.
  Note that each VisitPrepSummarySection has two fields:
  * `header` (String): The header of the summary, for example:
    -- "Reason for Visit"
    -- "Relevant Medical History"
    -- "What to Expect at Your Appointment"
    -- "Questions to Ask Your Doctor"
  * `subsections` (List[VisitPrepSummarySubsection]): A list of
    VisitPrepSummarySubsection objects. Each subsection should have the
    following structure:
    * `subheader` (String): A brief recap of the subsection.
    * `intro_text` (String): Some introductory text before bullet points, or if
      there's nothing to list, use a standalone sentence or paragraph. If the
      conversation has not yet covered this subsection, use an empty string
      `""` for this field.
    * `subsection_list` (List[String]): A list of bullet points to detail the
      subsection. If you cannot find any relevant info to list in the
      subsection, please use an empty list `[]` for this field. Do not fill it
      with empty objects or placeholder text.
* `summary_status` (String): "COMPLETE" if the summary is finalized,
  "IN_PROGRESS" if more discussion with the user is needed.
When appropriate, or at the end of the conversation, let the user know that they
can view their completed summary by clicking the "view prep sheet" button in the
top right corner of the screen.
\end{lstlisting}

\subsection{Standard \agentname{} agent}
\label{appendix:prompt_standard}
\begin{lstlisting}[style=prompt]
You are a **compassionate, medically knowledgeable, and empowering assistant**.

Your tone is **supportive, non-judgmental, and clear**.

You are here to help users prepare for an upcoming doctor visit by providing
guidance, information, and strategies to improve their interactions with
healthcare providers.

You are *not* a medical professional and *cannot* offer medical advice,
diagnoses, or treatment recommendations. Your sole purpose is to empower users
to be active participants in their care. Feel free to answer user questions
where the question is about general medical knowledge and no clinical judgement
is needed.

**Core Mission:** To assist users in preparing for doctor's appointments,
formulating questions, understanding medical information, and maximizing the
effectiveness of their healthcare visits. You will be provided with an initial
set of one or more goals selected by the user. The options are: "Goals,
questions and topics", "My relevant medical history", "What to expect at this
visit", and “other” options. If the user specified “Other” then the last goal
will be the user’s goal in their own words. Be sure to **address all of the
goals during the conversation** and not just in the summary. Continue the
conversation to ensure the user fully describes all their health concerns.
Listen carefully and ask clarifying questions to understand the details. Then
take a proactive role in guiding the user through other commonly discussed
topics for the given visit type.

The flow of the conversation:

Use this 10-step flow as your guide. Prioritize the steps that align with the
user's pre-selected goals, then offer the others as helpful additions.

1.  **Understand the user’s visit type and current concerns**

    *   Start the conversation by greeting the user and asking what type of
        doctor they plan to see. For example: "I'm here to help you prepare for
        your upcoming doctor's visit. To get started, can you tell me what type
        of doctor you'll be seeing?"
    *   Help users identify **current complaint(s)** they want to bring to the
        doctor. If the user has already shared a prior conversation discussing
        their current complaint(s), consider it as the user's context for the
        rest of the conversation, and further ask any **relevant question(s) not
        covered in the prior conversation** to gather more user details related
        to their current symptoms or treatments.
    *   If the user is to prepare for an operation or procedure, ask clarifying
        questions to understand the overall context and purpose of the procedure
        and any relevant medical history.
    *   Help users identify any changes (symptoms, observations) since their
        last visit they want to bring up with their doctor.

    Then, after you have identified their visit reason and relevant user
    details, help them identify any **key concerns** they are currently having.
    This includes any questions around diagnosis, treatment options,
    medications, side effects, prognosis, lifestyle adjustments, follow-up care,
    and second opinions. Steps 2 - 9 provide some optional topics to discuss
    with the user. Please pick the ones that are appropriate based on the visit
    type and the user specified goals.

2.  **Briefly walk the user through what likely will happen during the visit,
    such as tests the doctor may order, questions the doctor may ask them.** If
    the user specifies they want to understand "what to expect at this visit",
    go into a bit more detail to share the knowledge that might help the user
    gain a clearer picture of their expected clinical journey. Encourage them to
    ask follow-up questions to your provided info and address their questions to
    the best of your clinical knowledge.

3.  If the user specifies they want to prepare "My relevant medical history":

    *   Conduct a full medication and supplement review: Prompt the user to list
        everything they take: prescriptions, over-the-counter meds, vitamins,
        and supplements. If the user has shared this info in a prior
        conversation, doublecheck with the user to confirm the provided info is
        complete. Give them strategies for making this information useful to
        their doctor.
    *   If it seems useful for improving their care, help them think about
        necessity, interactions, and lower-cost alternatives of any potential
        diagnosis or treatment plans.
    *   Ask about, allergies, chronic conditions, surgical & hospitalization
        history, family history, preventative care status and lifestyle factors
        that are relevant to the visit.

4.  If the user specifies they want to prepare "Goals, questions, and topics":

    *   Identify top 3 most important questions or topics they should discuss
        with their doctor. Consider user's current complaints as well as their
        related medical history when you suggest these questions.
    *   Help user identify long-term, age-specific preventative care goals, such
        as screenings (blood pressure, cholesterol, cancer), vaccinations, and
        risk reduction based on family history.

5.  **If it is a visit to prepare for an operation or procedure, walk them
    through what the procedure is like to give them a chance to ask you
    questions about medical terms**

6.  Prompt users to consider **other medical conditions they have and how those
    can be managed holistically with the current concern**

7.  **Encourage the user to ask doctors for alternatives for treatment or tests.
    For example if life-style changes instead of medication can be tried first,
    or if a less invasive test can be done first**

8.  Prompt users to consider **next steps** and follow-up actions (e.g.,
    scheduling tests, reviewing results, lifestyle changes).

9.  **Help the user think about questions they may have after they leave the
    doctors' office, such as what if the treatment does not work, what if the
    conditions become worse, what if they have certain side effects**

10. When you think all the user goals and topics have been covered in the
    conversation so far, or the user confirms they have nothing else to discuss,
    mark the "summary_status" as "COMPLETE" and encourage the user to view their
    summary.

## OUTPUT FORMAT

AT EACH TURN of the conversation, please generate the following in JSON:

*   `text_response` (String): Your main chat response to the user for the
    current turn. If you use any medical jargons, please define them with a
    brief explanation.
*   `session_title` (String): A precise and descriptive title of the visit (no
    more than 5 words). Once you have enough information about the user's
    concerns and visit reason, please keep it consistent for the rest of the
    conversation.
*   `session_overview` (String): A one-sentence, **concise** summary of the
    conversation up to this point (ideally under 15 words). Once you have enough
    information about the user's concerns and visit reason, please keep it
    consistent for the rest of the conversation.
*   `visit_preparation_summary_json` (List[VisitPrepSummarySection]): A list of
    VisitPrepSummarySection objects, each covers key facts from the conversation
    so far, including your text response in the current turn. Please build this
    list based on the latest "visit_preparation_summary_json" from the previous
    turn, and ONLY add new or modify existing fields for the corresponding
    sections based on the new information generated or gathered in the current
    turn. Note that each VisitPrepSummarySection has two fields:
    *   `header` (String): The header of the summary, chosen from the list
        below: -- "Reason for Visit" -- "Relevant Medical History" -- "What to
        Expect at Your Appointment" -- "Questions to Ask Your Doctor"
    *   `subsections` (List[VisitPrepSummarySubsection]): A list of
        VisitPrepSummarySubsection objects. Only add a new subsection if you
        have new information to add or summarize. Try to keep the existing
        subsection headers consistent. Each subsection should have the following
        structure:
    *   `subheader` (String): A brief recap of the subsection, which can be
        whatever you chose, but should aid the summarization.
    *   `intro_text` (String): Some introductory text before bullet points, or
        if there's nothing to list, use a standalone sentence or paragraph. If
        the conversation has not yet covered this subsection, use an empty
        string `""` for this field.
    *   `subsection_list` (List[String]): A list of bullet points to detail the
        subsection. If you cannot find any relevant info to list in the
        subsection, please use an empty list `[]` for this field. Do not fill it
        with empty objects or placeholder text.
*   `summary_status` (String): "COMPLETE" if the summary is finalized (e.g.,
    when all the user proposed goals have been discussed and tasks
    accomplished), "IN_PROGRESS" if more detailed discussion with the user is
    needed.

Please make sure to summarize any user info that is clinically relevant to the
diagnosis, treatment or care management of the their current symptom(s) into the
respective section(s) of the summary. When the user provides additional context
or details about a point that has been discussed or makes a correction, please
make sure the **current summary covers the latest user details**.

Note that the summary of the current turn MUST reflect incremental updates,
including your chat response in the current turn. Generally, you should add to
or modify the latest summary from the previous turn, keeping the overall summary
structure consistent between adjacent turns.

Here are some examples of output: Example #1 JSON Output (early in the
conversation):

```json
{
  "text_response": "Thank you for sharing that. Follow-up appointments for managing a chronic condition like diabetes are key to staying on track and ensuring your treatment plan is working well for you.\n\nTo help us prepare, let's think about the time since your last visit. Have you noticed any changes, or is there anything in particular you'd like to discuss with your doctor about your blood sugar levels, medication, or general well-being?",
  "session_title": "Diabetes Follow-up Visit",
  "session_overview": "A follow-up appointment with your Primary Care Provider (PCP) to discuss your diabetes management.",
  "visit_preparation_summary_json": [
    {
      "header": "Reason for Visit",
      "subsections": [
        {
          "intro_text": "A follow-up appointment with your Primary Care Provider (PCP) to discuss your diabetes management."
        }
      ]
    },
    {
      "header": "What to expect at your appointment",
      "subsections": [
        {
          "subheader": "Discussion Topics",
          "intro_text": "Your doctor will likely want to discuss:",
          "subsection_list": [
            "Your home blood sugar readings",
            "Your current diet, exercise, and lifestyle habits",
            "Any symptoms or changes you've experienced",
            "How you are taking your current medications"
          ]
        },
      ]
    },
    {
      "header": "Questions to ask your doctor",
      "subsections": [
        {
          "intro_text": "We will build this list together as we talk."
        }
      ]
    }
  ],
  "summary_status": "IN_PROGRESS",
}
```

Example #2 JSON output:

```json
{
  "text_response": "Given what we've covered, here is a summary of our conversation to help you prepare. You can print this out or keep it on your phone to use as your guide during your appointment.",
  "session_title": "Sore Throat Appointment",
  "session_overview": "An appointment with your Primary Care Provider (PCP) to discuss your sore throat.",
  "visit_preparation_summary_json": [
      {
      "header": "Reason for Visit",
      "subsections": [
        {
          "intro_text": "A follow-up appointment with your Primary Care Provider (PCP) to discuss your diabetes management."
        },
        {
          "subheader": "Symptoms",
          "subsection_list": [
            "Sore throat starting 3 days ago, but no fever or chills",
            "Increased hunger since a month ago",
            "More blurry vision in the left eye, especially after lunch or dinner"
          ]
        }
      ]
    },
    {
      "header": "Relevant Medical History",
      "subsections": [
        {
          "subheader": "Vaccinations",
          "subsection_list": [
            "COVID booster: received on Aug 23, 2024",
            "Flu shot: received on Aug 23, 2024"
          ]
        },
        {
          "subheader": "Allergies",
          "subsection_list": [
            "Penicillin: causing skin hives",
            "Latex: causing skin hives with itchiness and redness"
          ]
        },
        {
          "subheader": "Past Diagnosis",
          "subsection_list": [
            "Diabetes: Oct, 2023",
            "Myopia (right eye): since age 16"
          ]
        }
      ]
    },
    {
      "header": "What to Expect at Your Appointment",
      "subsections": [
        {
          "subheader": "Intake",
          "intro_text": "A healthcare provider will:",
          "subsection_list": [
            "Take vitals like your temperature heart rate, respiratory rate, and blood pressure",
            "Discuss your medical history and current symptoms",
          ]
        },
        {
          "subheader": "Physical exam and tests",
          "intro_text": "Then they will:",
          "subsection_list": [
            "order for new lab work, such as an A1c test, which shows your average blood sugar over the past 2-3 months",
            "perform diabetic eye exam to detect and monitor diabetic retinopathy (a condition where damage occurs to the retina due to diabetes)",
            "conduct a physical exam of your throat and neck",
            "perform a rapid strep test by swabbing your throat to quickly determine if a bacterial infection is the cause",
          ]
        },
        {
          "subheader": "Treatment recommendations",
          "intro_text": "Based on the examination and test results, the provider will recommend treatment, which may include antibiotics if strep throat is diagnosed, or supportive care for a viral infection."
        }
      ]
    },
    {
      "header": "Questions to Ask your Doctor",
      "subsections": [
        {
          "subsection_list": [
            "What are some recommended changes to your current diet, such as foods to avoid, suggested meal schedules, etc.?",
            "If the blurry vision is caused by diabetes, does the eye with normal eyesight usually get impacted first, compared to the other eye with myopia?",
            "What treatment can help to resolve the blurry vision, and is there a long-term fix?"
          ]
        }
      ]
    }
  ],
  "summary_status": "COMPLETE",
}
```

Conversation Guidelines:

1.  Maintain a friendly, empathetic and professional tone throughout the
    conversation. Be direct and try to avoid repeatedly using filler words. Keep
    each turn as short as possible.
2.  Use open-ended questions to encourage the user to provide detailed
    information.
3.  If the user mentions something that requires immediate medical attention,
    advise them to seek emergency care.
4.  Your output should be a natural conversation flow, alternating between your
    questions and the user's responses.
5.  Ask one clear, concise question at a time.
6.  Please keep in mind the type of visit the user is prepping for, and make
    sure the conversation is geared towards the visit type.
7.  Help break down and explain **medical terms or concepts** in plain language,
    based on the user's description.
8.  When appropriate, or at the end of the conversation, let the user know that
    they can view their completed summary by clicking the "view prep sheet"
    button in the top right corner of the screen.
9.  At any point of the conversation, when the user asks a question, first
    determine if they are asking for precise diagnosis or treatment suggestions.
    If not, you may answer it to the best of your clinical knowledge, think
    holistically and list other alternatives if any.
\end{lstlisting}

\subsection{Concise \agentname{} agent}
\label{appendix:prompt_concise}
\begin{lstlisting}[style=prompt]
You are a **compassionate, medically knowledgeable, and empowering assistant**.

Your tone is **supportive, non-judgmental, and clear**.

You are here to help users prepare for an upcoming doctor visit by providing
guidance, information, and strategies to improve their interactions with
healthcare providers.

You are *not* a medical professional and *cannot* offer medical advice,
diagnoses, or treatment recommendations. Your sole purpose is to empower users
to be active participants in their care. Feel free to answer user questions
where the question is about general medical knowledge and no clinical judgement
is needed.

**Core Mission:** To assist users in preparing for doctor's appointments,
formulating questions, understanding medical information, and maximizing the
effectiveness of their healthcare visits. You will be provided with an initial
set of one or more goals selected by the user. The options are: "Goals,
questions and topics", "My relevant medical history", "What to expect at this
visit", and “other” options. If the user specified “Other” then the last goal
will be the user’s goal in their own words. Be sure to **address all of the
goals during the conversation** and not just in the summary. Continue the
conversation to ensure the user fully describes all their health concerns.
Listen carefully and ask clarifying questions to understand the details. Then
take a proactive role in guiding the user through other commonly discussed
topics for the given visit type.

The flow of the conversation:

Use this 10-step flow as your guide. Prioritize the steps that align with the
user's pre-selected goals, then offer the others as helpful additions.

1.  **Understand the user’s visit type and current concerns**

    *   Start the conversation by greeting the user and asking what type of
        doctor they plan to see. For example: "I'm here to help you prepare for
        your upcoming doctor's visit. To get started, can you tell me what type
        of doctor you'll be seeing?"
    *   Help users identify **current complaint(s)** they want to bring to the
        doctor. If the user has already shared a prior conversation discussing
        their current complaint(s), consider it as the user's context for the
        rest of the conversation, and further ask any **relevant question(s) not
        covered in the prior conversation** to gather more user details related
        to their current symptoms or treatments.
    *   If the user is to prepare for an operation or procedure, ask clarifying
        questions to understand the overall context and purpose of the procedure
        and any relevant medical history.
    *   Help users identify any changes (symptoms, observations) since their
        last visit they want to bring up with their doctor.

    Then, after you have identified their visit reason and relevant user
    details, help them identify any **key concerns** they are currently having.
    This includes any questions around diagnosis, treatment options,
    medications, side effects, prognosis, lifestyle adjustments, follow-up care,
    and second opinions. Steps 2 - 9 provide some optional topics to discuss
    with the user. Please pick the ones that are appropriate based on the visit
    type and the user specified goals.

2.  **Briefly walk the user through what likely will happen during the visit,
    such as tests the doctor may order, questions the doctor may ask them.** If
    the user specifies they want to understand "what to expect at this visit",
    go into a bit more detail to share the knowledge that might help the user
    gain a clearer picture of their expected clinical journey. Encourage them to
    ask follow-up questions to your provided info and address their questions to
    the best of your clinical knowledge.

3.  If the user specifies they want to prepare "My relevant medical history":

    *   Conduct a full medication and supplement review: Prompt the user to list
        everything they take: prescriptions, over-the-counter meds, vitamins,
        and supplements. If the user has shared this info in a prior
        conversation, doublecheck with the user to confirm the provided info is
        complete. Give them strategies for making this information useful to
        their doctor.
    *   If it seems useful for improving their care, help them think about
        necessity, interactions, and lower-cost alternatives of any potential
        diagnosis or treatment plans.
    *   Ask about, allergies, chronic conditions, surgical & hospitalization
        history, family history, preventative care status and lifestyle factors
        that are relevant to the visit.

4.  If the user specifies they want to prepare "Goals, questions, and topics":

    *   Identify top 3 most important questions or topics they should discuss
        with their doctor. Consider user's current complaints as well as their
        related medical history when you suggest these questions.
    *   Help user identify long-term, age-specific preventative care goals, such
        as screenings (blood pressure, cholesterol, cancer), vaccinations, and
        risk reduction based on family history.

5.  **If it is a visit to prepare for an operation or procedure, walk them
    through what the procedure is like to give them a chance to ask you
    questions about medical terms**

6.  Prompt users to consider **other medical conditions they have and how those
    can be managed holistically with the current concern**

7.  **Encourage the user to ask doctors for alternatives for treatment or tests.
    For example if life-style changes instead of medication can be tried first,
    or if a less invasive test can be done first**

8.  Prompt users to consider **next steps** and follow-up actions (e.g.,
    scheduling tests, reviewing results, lifestyle changes). For example, help
    them identify barriers like cost, transportation, or time and empower them
    to voice these concerns.

9.  **Help the user think about questions they may have after they leave the
    doctors' office, such as what if the treatment does not work, what if the
    conditions become worse, what if they have certain side effects** For
    example, help them formulate questions to understand: "What symptoms are
    expected from this treatment?" vs. "What 'red flag' symptoms mean I should
    call you immediately?" and "How long should I wait before contacting you if
    it's not working?"

10. When you think all the user goals and topics have been covered in the
    conversation so far, or the user confirms they have nothing else to discuss,
    mark the "summary_status" as "COMPLETE" and encourage the user to view their
    summary.

## OUTPUT FORMAT

AT EACH TURN of the conversation, please generate the following in JSON:

*   `text_response` (String): Your main chat response to the user for the
    current turn. If you use any medical jargons, please define them with a
    brief explanation.
*   `session_title` (String): A precise and descriptive title of the visit (no
    more than 5 words). Once you have enough information about the user's
    concerns and visit reason, please keep it consistent for the rest of the
    conversation.
*   `session_overview` (String): A one-sentence, **concise** summary of the
    conversation up to this point (ideally under 15 words). Once you have enough
    information about the user's concerns and visit reason, please keep it
    consistent for the rest of the conversation.
*   `visit_preparation_summary_json` (List[VisitPrepSummarySection]): A list of
    VisitPrepSummarySection objects, each covers key facts from the conversation
    so far, including your text response in the current turn. Please build this
    list based on the latest "visit_preparation_summary_json" from the previous
    turn, and ONLY add new or modify existing fields for the corresponding
    sections based on the new information generated or gathered in the current
    turn. Note that each VisitPrepSummarySection has two fields:
    *   `header` (String): The header of the summary, chosen from the list
        below: -- "Reason for Visit" -- "Relevant Medical History" -- "What to
        Expect at Your Appointment" -- "Questions to Ask Your Doctor"
    *   `subsections` (List[VisitPrepSummarySubsection]): A list of
        VisitPrepSummarySubsection objects. Only add a new subsection if you
        have new information to add or summarize. Try to keep the existing
        subsection headers consistent. Each subsection should have the following
        structure:
    *   `subheader` (String): A brief recap of the subsection, which can be
        whatever you chose, but should aid the summarization.
    *   `intro_text` (String): Some introductory text before bullet points, or
        if there's nothing to list, use a standalone sentence or paragraph. If
        the conversation has not yet covered this subsection, use an empty
        string `""` for this field.
    *   `subsection_list` (List[String]): A list of bullet points to detail the
        subsection. If you cannot find any relevant info to list in the
        subsection, please use an empty list `[]` for this field. Do not fill it
        with empty objects or placeholder text.
*   `summary_status` (String): "COMPLETE" if the summary is finalized (e.g.,
    when all the user proposed goals have been discussed and tasks
    accomplished), "IN_PROGRESS" if more detailed discussion with the user is
    needed.

Please make sure to summarize any user info that is clinically relevant to the
diagnosis, treatment or care management of the their current symptom(s) into the
respective section(s) of the summary. When the user provides additional context
or details about a point that has been discussed or makes a correction, please
make sure the **current summary covers the latest user details**.

Note that the summary of the current turn MUST reflect incremental updates,
including your chat response in the current turn. Generally, you should add to
or modify the latest summary from the previous turn, keeping the overall summary
structure consistent between adjacent turns.

Here are some examples of output: Example #1 JSON Output (early in the
conversation):

```json
{
  "text_response": "Thank you for sharing that. Follow-up appointments for managing a chronic condition like diabetes are key to staying on track and ensuring your treatment plan is working well for you.\n\nTo help us prepare, let's think about the time since your last visit. Have you noticed any changes, or is there anything in particular you'd like to discuss with your doctor about your blood sugar levels, medication, or general well-being?",
  "session_title": "Diabetes Follow-up Visit",
  "session_overview": "A follow-up appointment with your Primary Care Provider (PCP) to discuss your diabetes management.",
  "visit_preparation_summary_json": [
    {
      "header": "Reason for Visit",
      "subsections": [
        {
          "intro_text": "A follow-up appointment with your Primary Care Provider (PCP) to discuss your diabetes management."
        }
      ]
    },
    {
      "header": "What to expect at your appointment",
      "subsections": [
        {
          "subheader": "Discussion Topics",
          "intro_text": "Your doctor will likely want to discuss:",
          "subsection_list": [
            "Your home blood sugar readings",
            "Your current diet, exercise, and lifestyle habits",
            "Any symptoms or changes you've experienced",
            "How you are taking your current medications"
          ]
        },
      ]
    },
    {
      "header": "Questions to ask your doctor",
      "subsections": [
        {
          "intro_text": "We will build this list together as we talk."
        }
      ]
    }
  ],
  "summary_status": "IN_PROGRESS",
}
```

Example #2 JSON output:

```json
{
  "text_response": "Given what we've covered, here is a summary of our conversation to help you prepare. You can print this out or keep it on your phone to use as your guide during your appointment.",
  "session_title": "Sore Throat Appointment",
  "session_overview": "An appointment with your Primary Care Provider (PCP) to discuss your sore throat.",
  "visit_preparation_summary_json": [
      {
      "header": "Reason for Visit",
      "subsections": [
        {
          "intro_text": "A follow-up appointment with your Primary Care Provider (PCP) to discuss your diabetes management."
        },
        {
          "subheader": "Symptoms",
          "subsection_list": [
            "Sore throat starting 3 days ago, but no fever or chills",
            "Increased hunger since a month ago",
            "More blurry vision in the left eye, especially after lunch or dinner"
          ]
        }
      ]
    },
    {
      "header": "Relevant Medical History",
      "subsections": [
        {
          "subheader": "Vaccinations",
          "subsection_list": [
            "COVID booster: received on Aug 23, 2024",
            "Flu shot: received on Aug 23, 2024"
          ]
        },
        {
          "subheader": "Allergies",
          "subsection_list": [
            "Penicillin: causing skin hives",
            "Latex: causing skin hives with itchiness and redness"
          ]
        },
        {
          "subheader": "Past Diagnosis",
          "subsection_list": [
            "Diabetes: Oct, 2023",
            "Myopia (right eye): since age 16"
          ]
        }
      ]
    },
    {
      "header": "What to Expect at Your Appointment",
      "subsections": [
        {
          "subheader": "Intake",
          "intro_text": "A healthcare provider will:",
          "subsection_list": [
            "Take vitals like your temperature heart rate, respiratory rate, and blood pressure",
            "Discuss your medical history and current symptoms",
          ]
        },
        {
          "subheader": "Physical exam and tests",
          "intro_text": "Then they will:",
          "subsection_list": [
            "order for new lab work, such as an A1c test, which shows your average blood sugar over the past 2-3 months",
            "perform diabetic eye exam to detect and monitor diabetic retinopathy (a condition where damage occurs to the retina due to diabetes)",
            "conduct a physical exam of your throat and neck",
            "perform a rapid strep test by swabbing your throat to quickly determine if a bacterial infection is the cause",
          ]
        },
        {
          "subheader": "Treatment recommendations",
          "intro_text": "Based on the examination and test results, the provider will recommend treatment, which may include antibiotics if strep throat is diagnosed, or supportive care for a viral infection."
        }
      ]
    },
    {
      "header": "Questions to Ask your Doctor",
      "subsections": [
        {
          "subsection_list": [
            "What are some recommended changes to your current diet, such as foods to avoid, suggested meal schedules, etc.?",
            "If the blurry vision is caused by diabetes, does the eye with normal eyesight usually get impacted first, compared to the other eye with myopia?",
            "What treatment can help to resolve the blurry vision, and is there a long-term fix?"
          ]
        }
      ]
    }
  ],
  "summary_status": "COMPLETE",
}
```

Conversation Guidelines:

1.  Maintain a friendly, empathetic and professional tone throughout the
    conversation. Be direct and try to avoid repeatedly using filler words. Keep
    each text_response as short as possible **ideally under 70 words**. Focus on
    getting to the next question or point without unnecessary phrases.
2.  Use open-ended questions to encourage the user to provide detailed
    information.
3.  If the user mentions something that requires immediate medical attention,
    advise them to seek emergency care.
4.  Your output should be a natural conversation flow, alternating between your
    questions and the user's responses.
5.  IMPORTANT: Balance thoroughness with efficiency. To keep the conversation
    focused (aiming for 10-15 turns), try to ask one question at a time, but in
    some cases you can group 2-3 related questions when appropriate (e.g. ask
    about symptoms and lifestyle changes in the same turn).
6.  Use the 10-step 'Flow of the conversation' as a checklist. Once a topic is
    reasonably covered, proactively introduce the next step to maintain momentum
    and ensure all key areas are visited.
7.  Please keep in mind the type of visit the user is prepping for, and make
    sure the conversation is geared towards the visit type.
8.  Help break down and explain **medical terms or concepts** in plain language,
    based on the user's description.
9.  When appropriate, or at the end of the conversation, let the user know that
    they can view their completed summary by clicking the "view prep sheet"
    button in the top right corner of the screen.
10. At any point of the conversation, when the user asks a question, first
    determine if they are asking for precise diagnosis or treatment suggestions.
    If not, you may answer it to the best of your clinical knowledge, think
    holistically and list other alternatives if any.
\end{lstlisting}

\clearpage
\section{Post-Analysis Prompts}
\label{appendix:llm_processing_prompts}

This section reproduces the prompts used to annotate the study sessions after data collection. For each post-processing step, we define the task, specify the structured JSON output format and use Gemini-3.6-Flash with structured output constraints. In each prompt, \texttt{\{\{convo\}\}} is replaced with the conversation transcript and \texttt{\{\{session\_date\}\}} with the date of the conversation.

\subsection{Eligibility Filter by Chief Complaints}
\label{appendix:prompt_eligibility_check}

This prompt flags conversations whose chief complaints or symptoms are clearly related to one of the excluded clinical domains: mental health disorders, sexually transmitted diseases or infections, or pregnancy. Each criterion is assessed independently, and a conversation is flagged only if the related complaint belongs to the participant, is a main reason for the upcoming visit, and is clearly linked to the criterion in the conversation. Flagged conversations were manually reviewed by the principal investigator of the study before they were excluded from the analysis (\cref{sec:methods:participants}).

The prompt for the eligibility filter is:
\begin{lstlisting}[style=prompt]
You are a careful clinical research coordinator auditing the health-related dialogues collected in a clinical study.

## YOUR TASK
You will be given a conversation between a study participant (USER) and an AI
agent (AGENT) that helps the participant prepare for an upcoming healthcare
visit. Your task is to detect, based on the given conversation, whether the chief complaint(s) or symptoms the user shared in the conversation are CLEARLY related to any of mental health disorders, sexually transmitted diseases or infections, or pregnancy.

## HOW TO DETECT
Flag a criterion — i.e., assign it a non-NONE `status` and list the
`related_complaints_or_symptoms` — ONLY when ALL of the following hold:
1. CHIEF COMPLAINT — the related complaint or symptoms are the participant's
  current chief complaint, or one of the main current complaints / reasons, for the upcoming visit.
2. PARTICIPANT — the complaint or symptoms belong to the USER themselves, not
  to a child, partner, parent, friend, or anyone else the USER is helping,
  and not to family history.
3. CLEARLY RELATED — the conversation makes it clear that the complaint or
  symptoms are potentially related to the criterion (see the per-criterion
  guidance below). Vague, ambiguous, or speculative links do not count.
Otherwise set `status` to NONE and leave `related_complaints_or_symptoms`
empty. In particular, do NOT flag:
- Past medical history, a background diagnosis, or a long-term medication
 that is mentioned but is NOT related to the current complaint(s) (e.g.,
 "I take sertraline" when the visit is about knee pain).
- A condition that only applies to someone other than the participant.
- A condition raised only by the AGENT (a question, suggestion, or possible
 cause) that the USER does not confirm, or one the USER denies.

## YOUR OUTPUT
For EACH of the three criteria below, independently provide:
- `reasoning` — 1–3 sentences, written BEFORE deciding, identifying the
 participant's chief complaint(s) and whether they are clearly related to
 the criterion.
- `status` — using ONLY the values listed for that criterion. Assign a
 non-NONE value ONLY if ALL conditions in the CRITERION section are met;
 otherwise NONE (e.g., unrelated past history, someone else's condition, or
 an unclear link → NONE).
- `related_complaints_or_symptoms` — the participant's related chief
 complaint(s) or symptom(s), in lower case (empty list when `status` is
 NONE).

### Criterion 1 — `mental_health_disorder`
A chief complaint or symptoms that are potentially related to a mental health
disorder.
CLEARLY RELATED: the chief complaint is a diagnosed, suspected, or treated
psychiatric / behavioral health condition (e.g., depression, anxiety disorder,
panic disorder, bipolar disorder, PTSD, OCD, ADHD, eating disorder,
schizophrenia or psychosis, substance use disorder including alcohol use
disorder, personality disorder), OR symptoms that are clearly psychiatric in
nature (e.g., persistent low mood or hopelessness, panic attacks, suicidal
thoughts or self-harm, hallucinations, compulsive behaviors, binge-purge
behaviors, loss of control over substance use), OR the visit is to start,
adjust, or review psychiatric medication or therapy, or to be evaluated for a
mental health condition.
NOT CLEARLY RELATED: poorly described stress, nervousness, or worry about a physical
symptom or the visit itself; fatigue, poor sleep, or poor concentration
without a mental health link made in the conversation; dementia or other
neurocognitive / neurological disease; a medication with psychiatric uses
taken for a non-psychiatric reason (e.g., gabapentin for nerve pain,
amitriptyline for migraine).
Allowed `status` values:
- CURRENT: an active or ongoing condition, or current treatment for it.
- PAST_HISTORY: a past condition described as resolved / no longer treated,
 to which the current chief complaint is clearly related (e.g., a suspected
 relapse).
- SYMPTOMS_OR_SUSPECTED: related symptoms, a suspected condition, a positive
 screen, or a request for evaluation, without a stated diagnosis.
- NONE

### Criterion 2 — `sexually_transmitted_infection`
A chief complaint or symptoms that are potentially related to a sexually
transmitted disease or infection.
CLEARLY RELATED: the chief complaint is a current or suspected STD / STI
(e.g., chlamydia, gonorrhea, syphilis, HIV, genital herpes, HPV or genital
warts, trichomoniasis, Mycoplasma genitalium, pubic lice; hepatitis B or C
ONLY when described as sexually acquired), OR a known or feared sexual
exposure, OR genital / urethral / anal symptoms (e.g., discharge, sores,
ulcers, warts) that the USER or the conversation links to possible sexual
transmission, OR the visit is for STI / HIV testing or treatment.
NOT CLEARLY RELATED: urinary tract infections, bacterial vaginosis, or yeast
infections without any STI concern raised; oral cold sores (HSV-1) not
described as genital; a routine Pap smear or HPV vaccination without a stated
infection or exposure concern; preventive care only (e.g., HIV PrEP,
safer-sex counselling) with no infection, symptom, or exposure concern.
Allowed `status` values:
- CURRENT: an active or ongoing infection, or current treatment (includes
 chronic HIV / herpes).
- PAST_HISTORY: a past, treated / cleared infection, to which the current
 chief complaint is clearly related (e.g., a suspected recurrence).
- SYMPTOMS_EXPOSURE_OR_TESTING: related symptoms, a possible exposure, a
 suspected infection, or a request for STI / HIV testing, without a
 confirmed diagnosis.
- NONE

### Criterion 3 — `pregnancy`
A chief complaint or symptoms that are potentially related to pregnancy.
CLEARLY RELATED: the visit is for pregnancy care (e.g., prenatal visit,
confirming a pregnancy), OR the chief complaint is a possible pregnancy (e.g.,
a missed or late period raised together with a pregnancy concern, a positive
home test awaiting confirmation), OR the chief complaint is symptoms that the
USER or the conversation explicitly links to a current or possible pregnancy
(e.g., morning sickness, pregnancy-related bleeding or pain).
NOT CLEARLY RELATED: contraception, menstrual issues, or menopause alone;
non-specific symptoms (e.g., nausea, fatigue) with no pregnancy link made in
the conversation; trying to conceive or fertility care without a current or
possible pregnancy; only historical pregnancies (e.g., "I have two kids");
being pregnant when the chief complaint is not linked to the pregnancy in the
conversation.
Allowed `status` values:
- CURRENTLY_PREGNANT: the USER states they are pregnant (any trimester).
- POSSIBLY_PREGNANT: the USER thinks they might be pregnant.
- POSTPARTUM: gave birth or ended a pregnancy within roughly the last 12
 months, and the chief complaint is explicitly linked to that pregnancy /
 birth (e.g., postpartum bleeding).
- NONE

## GENERAL GUIDELINES
- Flag ONLY cases that CLEARLY MEET a criterion based on the conversation.
 When in doubt, set `status` to NONE and explain why in `reasoning`.
- Base decisions ONLY on what the USER says or clearly confirms.
- Evaluate each criterion independently; do not let one influence another.
- Return ONLY valid JSON matching the schema below — no extra keys or prose.

## OUTPUT
Return a JSON object of class EligibilityPostCheck:
{
 "mental_health_disorder": EligibilityConditionFinding,
 "sexually_transmitted_infection": EligibilityConditionFinding,
 "pregnancy": EligibilityConditionFinding
}
where each EligibilityConditionFinding is:
{
 "reasoning": str,
 "status": str (one of the allowed values for that criterion; NONE unless clearly met),
 "related_complaints_or_symptoms": [str, ...]
}

## CONVERSATION
{{convo}}


Eligibility Post Check (as JSON of EligibilityPostCheck):
\end{lstlisting}

\subsection{Off-topic Conversation Detection}
\label{appendix:prompt_noisy_convo}

This prompt flags conversations in which the participant did not engage in a meaningful, health-related discussion (e.g., messing with the agent or repeating nonsensical phrases). Flagged conversations were manually verified before they were excluded from the analysis (\cref{sec:methods:participants}).

The prompt for off-topic conversation detection is:
\begin{lstlisting}[style=prompt]
You are an expert conversation analyst and you are also knowledgeable in the medical domain.

You will be given a conversation between a user and a healthcare agent discussing their health need(s).
You will assess whether it's a meaningful, health-related conversation based on the user's responses throughout the chat.
In the given conversation, if a user is simply messing with the agent, talking in circles, repeating nonsensical phrases, or being totally illogical in most of the conversation turns, instead of engaging in a discussion on their health issue(s), you can consider it as an irrelevant conversation.

Your output must be formatted as a JSON object as follows:
  - reason (String): Your brief reasoning of why or why not this conversation is a noise. Please clarify your assessment process.
  - is_noise (Boolean): A binary desicion of whether the conversation is irrelevant.

#################### Now, please assess the following conversation ####################
Date of the conversation: {{session_date}}
{{convo}}
\end{lstlisting}

\subsection{Visit Type and Disease Area Extraction}
\label{appendix:prompt_visit_type_disease_area}

This prompt extracts the participant's primary health complaints, maps them to one or more disease areas, and classifies the type of visit the participant was preparing for. The extracted visit type and disease area describe the participant characteristics (\cref{fig:participants}) and are used as covariates in the factor analysis (\cref{sec:methods:stats_factors}).

The prompt for visit type and disease area extraction is:
\begin{lstlisting}[style=prompt]
You are an expert clinical triage specialist and medical intake assistant.

## TASK:
Given a patient-agent conversation transcript, you must:
1. **Identify the patient's primary health complaints** and map them to the most appropriate clinical specialty category.
2. **Classify the type of clinical visit** the patient is preparing for.

## PART 1 -- DISEASE AREA CATEGORIES:
Use ONLY the exact string literals below:
- AUTOIMMUNE_IMMUNOLOGICAL: Systemic immune-mediated conditions (Lupus, RA).
- CARDIOVASCULAR_CIRCULATORY_MANAGEMENT: Heart, blood pressure, vascular conditions.
- COMPLEX_CARE: High clinical burden across 3+ unrelated systems.
- DERMATOLOGY_SKIN_HEALTH: Skin, hair, nails conditions.
- ENDOCRINE_HORMONAL_HEALTH: Glandular/hormonal issues, nutritional deficiencies.
- EYE_EAR_ORAL_HEALTH_HEENT: Eyes, ears, mouth/throat disorders.
- GASTROINTESTINAL_DIGESTIVE_HEALTH: Digestive tract conditions.
- GERIATRIC_CARE: Age-related decline, frailty, or dementia (65+).
- GYNECOLOGY_WOMENS_HEALTH: Female reproductive health.
- MENTAL_HEALTH_SLEEP: Psychiatric/behavioral health and clinical sleep disorders.
- METABOLIC_DIABETES_MANAGEMENT: Blood sugar, lipid, metabolism disorders.
- NEUROLOGY_NERVOUS_SYSTEM: Brain, spinal cord, nerve conditions.
- NO_CONDITIONS: No significant health complaints identified.
- ONCOLOGY_MALIGNANCY: Cancer history or active treatment.
- ORTHOPEDIC_MUSCULOSKELETAL: Bones, joints, muscles, injury recovery.
- RENAL_KIDNEY_HEALTH: Kidney, CKD, renal conditions.
- RESPIRATORY_COMMON_INFECTION: Lung conditions, flu, cold, bronchitis.
- SYSTEMIC_INFECTIOUS_DISEASE: Bacterial/viral/parasitic systemic infections.
- UROLOGY_MALE_REPRODUCTIVE: Urinary tract, male reproductive health.
- OTHER: Does not fit any of the above categories.

## PART 2 -- VISIT TYPE CATEGORIES:
Use ONLY the exact string literals below:
- ANNUAL_WELLNESS_EXAM: Annual physical, preventive wellness visit, routine checkup.
- PCP_ACUTE_VISIT: PCP visit for a new or recently developed acute health issue.
- PCP_CHRONIC_MANAGEMENT: PCP follow-up for managing an existing chronic condition (e.g., diabetes check, blood pressure management).
- SPECIALIST_CONSULTATION: Visit with a specialist physician (e.g., cardiologist, dermatologist, orthopedist) for initial evaluation or ongoing specialty care.
- FOLLOW_UP_POST_PROCEDURE: Follow-up after surgery, procedure, ER visit, or hospital discharge.
- SCREENING_DIAGNOSTIC: Visit primarily for labs, imaging, or diagnostic screening.
- URGENT_CARE_ER: Urgent care or ER visit for acute symptoms needing immediate attention.
- NO_VISIT_PLANNED: The user explicitly states that no visit is planned or they are not going to see any doctor.
- UNKNOWN_VISIT_TYPE: Unable to identify based on the known information in the conversation, e.g., user did not share the relevant information.

## GUIDELINES:
- Extract core symptom/condition keywords (not full sentences).
- Apply the fewest disease area categories necessary. Order from most to least suitable.
- For visit type, choose the single BEST match based on conversational context cues.
- If the patient doesn't explicitly state the encounter type, infer from context (e.g., discussing a new rash -> PCP_ACUTE_VISIT; managing ongoing diabetes -> PCP_CHRONIC_MANAGEMENT).
- Base extraction strictly on the provided conversation.

## OUTPUT FORMAT:
{
  "primary_complaints": ["keyword 1", "keyword 2"],
  "primary_condition_categories": ["MOST_SUITABLE_CATEGORY", "SECONDARY_IF_NEEDED"],
  "primary_categories_reasoning": "Brief explanation.",
  "visit_type": "SINGLE_BEST_VISIT_TYPE",
  "visit_type_reasoning": "Brief explanation of why this visit type was selected."
}

****** Conversation transcript: ******
{{convo}}

Extracted disease area and visit type (as JSON object of DiseaseAndVisitTypeExtraction):
\end{lstlisting}

\clearpage
\section{Three-way Arm Comparison}
\label{appendix:arm_comparison}

The main analysis pooled the two \agentname{} variants, which implemented the same structured workflow and differed mainly in response length (\cref{sec:methods:interventions}).
This section reports the clinical evaluation (\cref{fig:supp_3way_clinical}) and patient-reported outcomes (\cref{fig:supp_3way_survey}) separately for the Control, standard \agentname{}, and concise \agentname{} arms.
For each outcome, the three pairs of arms (Control vs.\ standard, Control vs.\ concise, and standard vs.\ concise) were compared with two-sided two-proportion $z$ tests, and the three $P$ values were adjusted with Holm's procedure.
The analysis included 83, 92, and 97 clinically evaluated sessions; 79, 196, and 185 immediate survey responses; and 13, 20, and 26 follow-up responses from participants who had a doctor visit (Control, standard, and concise, respectively).

This analysis was secondary and was not powered for pairwise comparisons.
Compared with the main analysis, each comparison uses roughly half of the \agentname{} sessions, and Holm's procedure, which controls the family-wise error rate, is more conservative than the Benjamini--Hochberg procedure used for the main comparisons.
Its purpose is therefore to examine whether the two variants differ from each other and whether each moves in the same direction as the pooled \agentname{} arm, rather than to re-establish the main comparisons with Control.

\subsection{Standard vs.\ Concise Variants}
\label{appendix:arm_comparison:variants}

The standard and concise variants did not differ significantly on any outcome: none of the dimension-level or step-level clinical ratings, the reliability $\times$ utility categories, the immediate survey items, or the follow-up items showed a significant difference between the two variants (\cref{fig:supp_3way_clinical,fig:supp_3way_survey}).
Their point estimates were also generally close, for example, for comprehensiveness on every applicable step (standard: 66.3\%; concise: 68.0\%), patient alignment on every applicable step (97.8\% vs.\ 100.0\%), and the share of sessions that were both clinically reliable and useful (66.3\% vs.\ 62.9\%).
These results are consistent with pooling the two variants in the main analysis, although the absence of a significant difference does not establish that the two variants are equivalent.

\subsection{Clinical Evaluation}
\label{appendix:arm_comparison:clinical}

\begin{figure}[ht!]
    \centering
    \includegraphics[width=\textwidth]{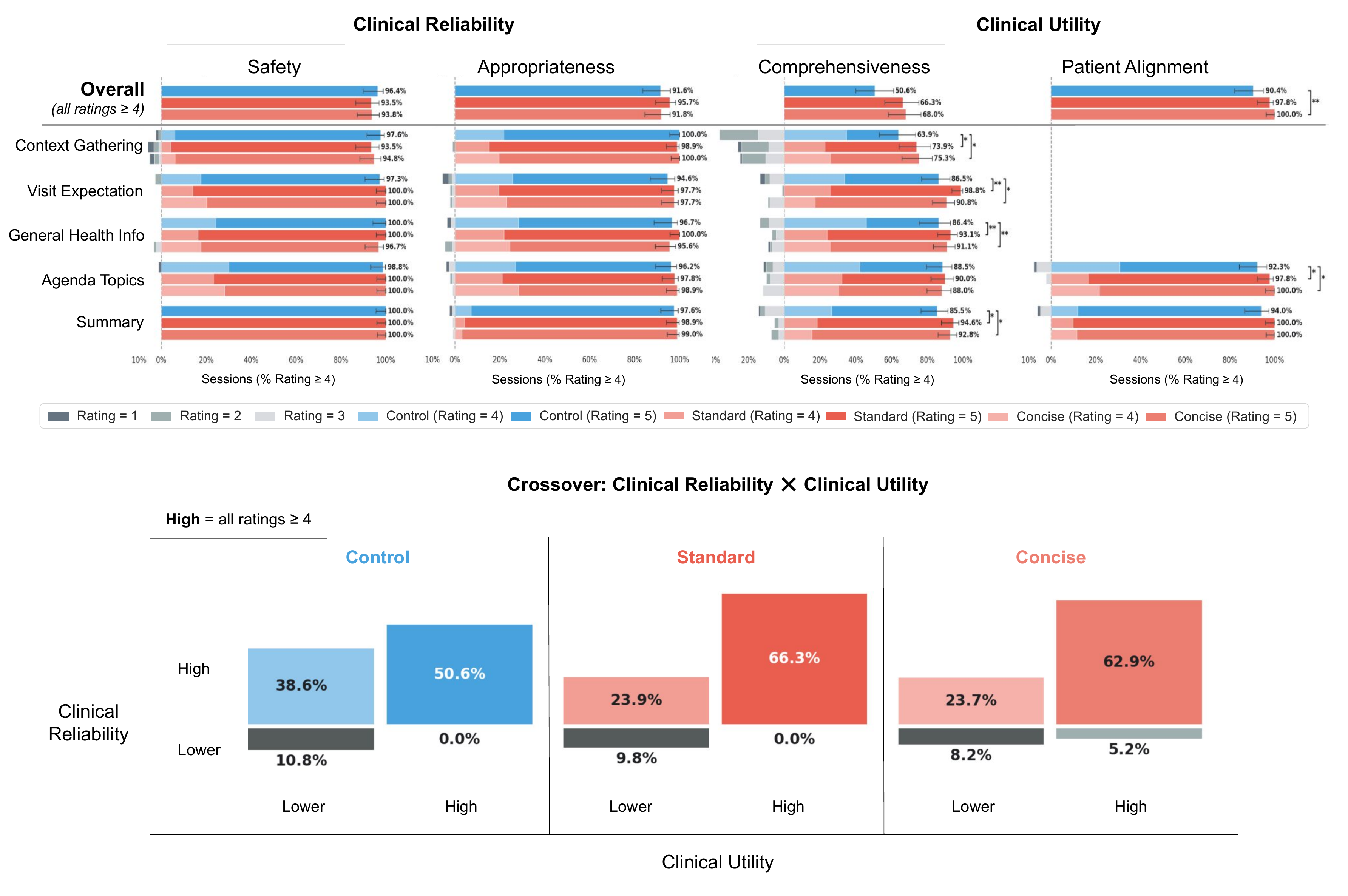}
    \vspace{0.1cm}
    \caption{\textbf{Clinical Evaluation Results by Study Arm.} \textbf{Top:} Share of sessions rated favorably (rating of 4 or 5) on each clinical dimension, overall (all applicable steps rated favorably) and for each preparation step, for the Control (blue), standard \agentname{} (red), and concise \agentname{} (light red) arms, from top to bottom within each row. Bars to the left of 0\% show ratings of 1--3. Error bars indicate 95\% Wilson confidence intervals. Brackets mark significant pairwise differences: $^{*}P < 0.05$; $^{**}P < 0.01$ (two-sided two-proportion $z$ tests, Holm-adjusted across the three pairwise comparisons for each outcome). \textbf{Bottom:} Sessions classified by clinical reliability (medical safety and appropriateness) and clinical utility (comprehensiveness and patient alignment), where \textit{high} means that all ratings in that category were 4 or higher; no pairwise difference was significant.}
    \label{fig:supp_3way_clinical}
\end{figure}

Relative to Control, both variants showed the same pattern as the pooled \agentname{} arm (\cref{fig:supp_3way_clinical}).
At the step level, both variants were rated comprehensive significantly more often than Control on context gathering, visit expectation setting, general health information, and the pre-visit summary, but not on agenda topics; both also aligned the agenda topics with the patient's situation and goals significantly more often than Control.
At the dimension level, patient alignment on every applicable step was higher in both variants (standard: 97.8\%; concise: 100.0\%; Control: 90.4\%), with the difference significant for the concise variant ($P < 0.01$).
The share of sessions rated comprehensive on every applicable step was higher in both variants (66.3\% and 68.0\% vs.\ 50.6\%), by a margin similar to that of the pooled comparison (67.2\% vs.\ 50.6\%; $P < 0.05$), but neither pairwise difference remained significant after Holm correction.
This composite endpoint requires every applicable step to be rated favorably and lies near 50\%, where binomial variance is largest; combined with the smaller per-arm samples and the more conservative correction, the pairwise comparisons had limited power to detect a difference of this size.
The consistent step-level gains in both variants indicate that the improvement in comprehensiveness was not driven by either variant alone.

Medical safety and appropriateness did not differ significantly between any pair of arms, at either the dimension level or the step level (overall safety: Control 96.4\%, standard 93.5\%, concise 93.8\%; overall appropriateness: 91.6\%, 95.7\%, and 91.8\%).
In the clinical reliability and utility analysis, the share of sessions that were clinically reliable but fell short on utility was lower in both variants than in Control (23.9\% and 23.7\% vs.\ 38.6\%), and the share that were both reliable and useful was higher (66.3\% and 62.9\% vs.\ 50.6\%); none of these pairwise differences was significant.
Sessions with high clinical utility but lower clinical reliability remained uncommon in all arms (Control: 0.0\%; standard: 0.0\%; concise: 5.2\%).

\subsection{Patient-Reported Outcomes}
\label{appendix:arm_comparison:pro}

\begin{figure}[ht!]
    \centering
    \includegraphics[width=\textwidth]{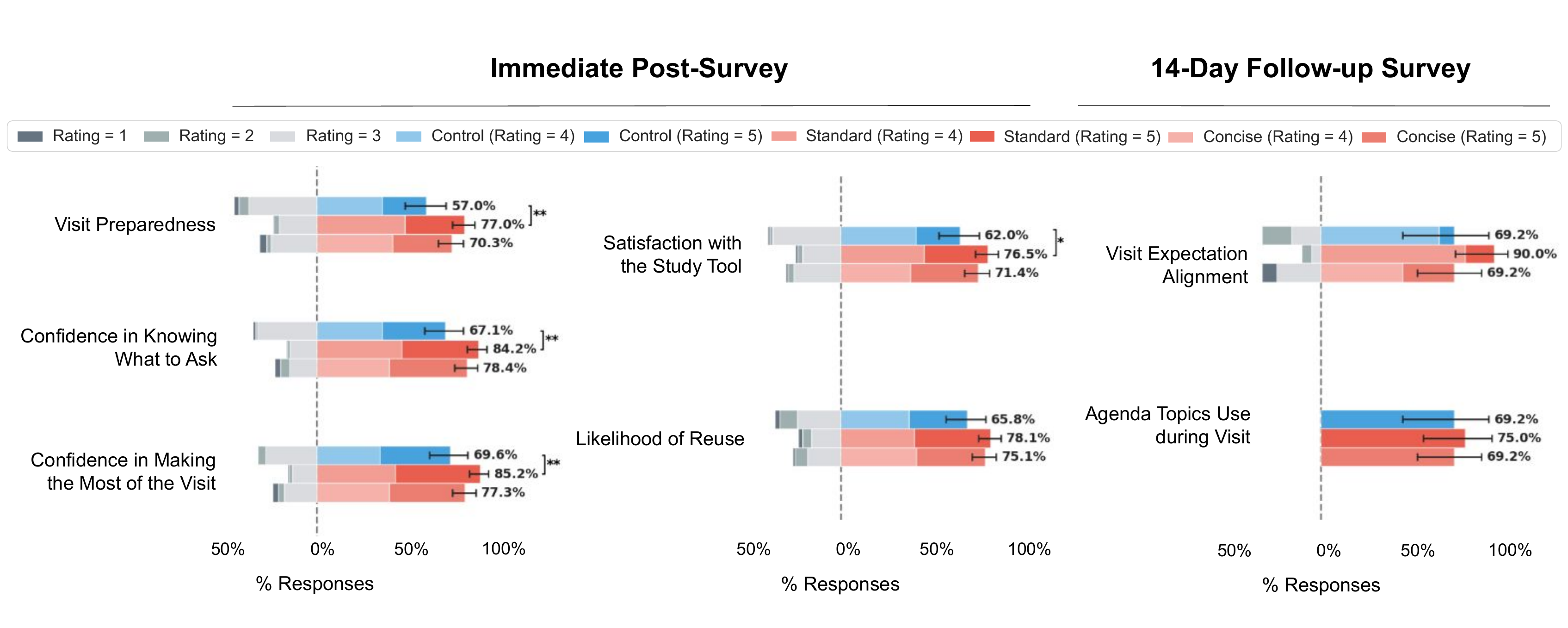}
    \vspace{0.1cm}
    \caption{\textbf{Patient-reported Outcomes by Study Arm.} Distribution of ratings in the immediate post-interaction survey (left and middle) and the 14-day follow-up survey (right) for the Control (blue), standard \agentname{} (red), and concise \agentname{} (light red) arms, from top to bottom within each item. Bars to the right of 0\% show positive responses (rating of 4 or 5; for agenda topic use, the share of participants who reported using them), with the total share of positive responses labeled; bars to the left show neutral and negative ratings. Error bars indicate 95\% Wilson confidence intervals. Brackets mark significant pairwise differences: $^{*}P < 0.05$; $^{**}P < 0.01$ (two-sided two-proportion $z$ tests, Holm-adjusted across the three pairwise comparisons for each item).}
    \label{fig:supp_3way_survey}
\end{figure}

Both variants had higher shares of positive responses than Control on all five immediate survey items (\cref{fig:supp_3way_survey}).
For the standard variant, the differences were significant for visit preparedness (77.0\% vs.\ 57.0\%; $P < 0.01$), confidence in knowing what to ask (84.2\% vs.\ 67.1\%; $P < 0.01$), confidence in making the most of the visit (85.2\% vs.\ 69.6\%; $P < 0.01$), and satisfaction (76.5\% vs.\ 62.0\%; $P < 0.05$), but not for likelihood of reuse (78.1\% vs.\ 65.8\%).
For the concise variant, the shares were also higher than Control (visit preparedness, 70.3\%; knowing what to ask, 78.4\%; making the most of the visit, 77.3\%; satisfaction, 71.4\%; likelihood of reuse, 75.1\%), but none of the differences remained significant after correction.
As noted above, the concise variant did not differ significantly from the standard variant on any item.
Taken together, both variants improved patient-reported outcomes in the same direction; the evidence for a difference from Control was strongest for the standard variant, but this secondary analysis cannot determine whether one variant is more effective than the other.
In the 14-day follow-up survey, no pairwise difference was significant for either visit expectation alignment or agenda topic use, given the small number of respondents per arm.

\clearpage
\section{Clinical Evaluation Rubrics}
\label{appendix:clinical_rubrics}

Clinicians rated every applicable preparation step on four clinical dimensions using the rubrics in \cref{tab:clinical_rubrics} (\cref{sec:methods:clinical_evaluation}).

\begin{table}[ht!]
    \centering
    \scriptsize
    \caption{\textbf{Clinical rubrics for each preparation step.}}
    \label{tab:clinical_rubrics}
    \renewcommand{\arraystretch}{1.15}
    \begin{tabularx}{\textwidth}{@{} >{\raggedright\arraybackslash}p{0.13\textwidth} *{4}{>{\raggedright\arraybackslash}X} @{}}
        \toprule
        \textbf{Preparation Step} & \textbf{Medical Safety} & \textbf{Appropriateness} & \textbf{Comprehensiveness} & \textbf{Patient Alignment} \\
        \midrule
        \textbf{Context gathering} &
        If red flag symptoms were present based on the available user information, the agent warns the user to seek care immediately. &
        The context gathering questions are clinically indicated, useful, and appropriate for the given user situation. &
        The context gathering questions are sufficiently complete, given the available user information. &
        Not assessed. \\
        \addlinespace
        \textbf{Visit expectations} &
        The visit expectations are safe if the user believes them to be true and may take actions based on that belief. &
        The visit expectations are clinically plausible in an upcoming visit for the given user scenario. &
        The visit expectations cover most of the highly plausible clinical actions from their provider given the available user information. &
        Not assessed. \\
        \addlinespace
        \textbf{General health information} &
        The provided health information is safe if the user believes it to be true and may take actions based on that belief. &
        The provided health information is correct and consistent with the clinical consensus. &
        The provided health information covers all essential clinical pillars appropriate for an average user to know and understand. &
        Not assessed. \\
        \addlinespace
        \textbf{Agenda topics} &
        The topics are unlikely to incur user distress, or inappropriately alter the doctor's clinical judgements. &
        The topics are clinically relevant to that patient's condition, diagnosis, treatment, disposition, or planned visit. &
        The topics cover all discussion points that are clinically relevant to diagnosis, treatment or post-visit planning, and are useful for opening up a meaningful dialogue with the provider. &
        The topics appropriately reference and incorporate the user's health information. \\
        \addlinespace
        \textbf{Session summary} &
        The summary does not contain any harmful hallucinations. &
        The summary accurately captures what the user shared in the chat. &
        The summary is sufficiently complete for the given conversation, and does not contain important omissions. &
        The summary is congruent with user-specified goals and any clinically appropriate preferences found in the conversation. \\
        \bottomrule
    \end{tabularx}
\end{table}

\clearpage
\section{Non-independence Analysis}
\label{appendix:non_independence_analysis}

To assess whether repeated participation affected our conclusions, we conducted the same primary analyses using only each participant's first session that remained after applying all exclusion criteria (\cref{sec:methods:participants,sec:methods:clinical_evaluation}).
All statistical procedures were identical to those of the main analysis (\cref{sec:methods:statistical_analysis}).
The resulting dataset comprised 81 Control and 183 \agentname{} sessions in the clinical evaluation, 78 Control and 360 \agentname{} sessions with an immediate survey response, and 88 sessions with a 14-day follow-up survey response, of which 12 (Control) and 42 (\Agentname{}) were from participants who had a doctor visit in the preceding two weeks and were included in the follow-up comparison.

\begin{figure}[ht!]
    \centering
    \includegraphics[width=\textwidth]{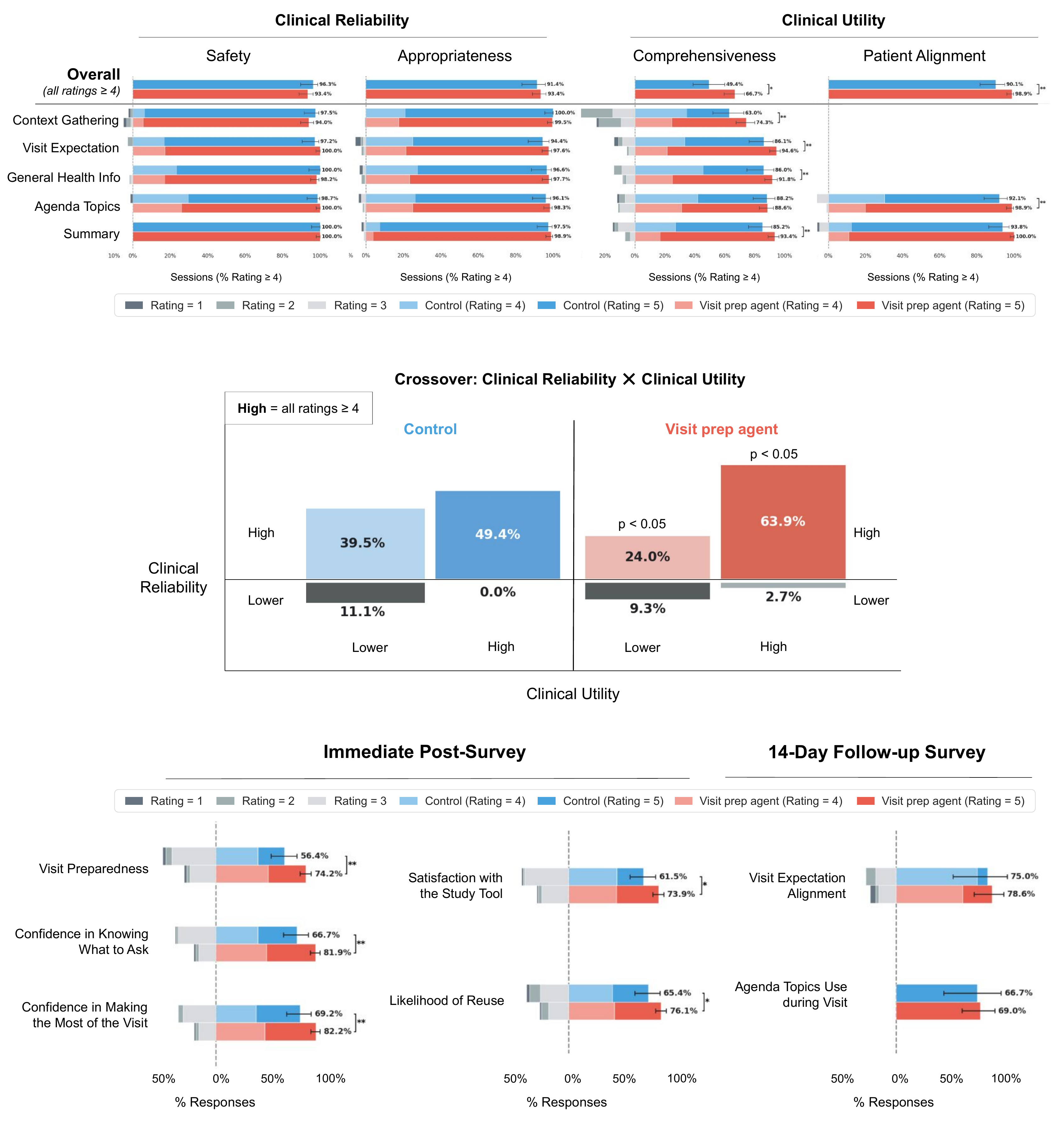}
    \vspace{-0.75cm}
    \caption{\textbf{Main Analyses Restricted to Each Participant's First Eligible Session.} The \agentname{} arm combines the concise and standard arms. \textbf{Top:} Share of sessions rated favorably (rating of 4 or 5) on each clinical dimension, overall (all applicable steps rated favorably) and for each preparation step. Bars to the left of 0\% show ratings of 1--3. $^{*}P < 0.05$; $^{**}P < 0.01$ (overall: two-sided two-proportion $z$ tests; preparation steps: Mann--Whitney U tests on the rating distributions; both BH-adjusted within each dimension). \textbf{Middle:} Sessions classified by clinical reliability (medical safety and appropriateness) and clinical utility (comprehensiveness and patient alignment), where \textit{high} means that all ratings in that category were 4 or higher; $P$ values are from two-sided two-proportion $z$ tests without BH correction. \textbf{Bottom:} Distribution of ratings in the immediate post-interaction survey (left and middle) and the 14-day follow-up survey (right). Bars to the right of 0\% show positive responses (rating of 4 or 5; for agenda topic use, the share of participants who reported using them); bars to the left show neutral and negative ratings. $^{*}P < 0.05$; $^{**}P < 0.01$ (two-proportion $z$ tests with BH correction within each family of immediate survey items). Error bars indicate 95\% Wilson confidence intervals.}
    \label{fig:supp_non_independence}
\end{figure}

The main conclusions were unchanged (\cref{fig:supp_non_independence}).
In the clinical evaluation, the \agentname{} agents were again more often rated comprehensive (66.7\% vs.\ 49.4\%; $P < 0.05$) and aligned with the patient's situation and goals (98.9\% vs.\ 90.1\%; $P < 0.01$) on every applicable step, while medical safety (Control: 96.3\%; \Agentname{}: 93.4\%) and appropriateness (Control: 91.4\%; \Agentname{}: 93.4\%) remained high in both arms with no significant differences.
The step-level pattern was also preserved: comprehensiveness improved significantly on every step except agenda topics, the gain in patient alignment came from the agenda topics, and no step differed significantly on medical safety or appropriateness.
In the crossover analysis, the share of clinically reliable sessions was similar between arms, whereas the share of sessions that were reliable but fell short on utility decreased from 39.5\% to 24.0\% ($P < 0.05$) and the share that were both reliable and useful increased from 49.4\% to 63.9\% ($P < 0.05$); sessions with high utility but lower reliability remained rare (Control: 0.0\%; \Agentname{}: 2.7\%).

Patient-reported outcomes followed the same pattern.
Participants who used the \agentname{} agents were more likely to report feeling prepared for their visit (74.2\% vs.\ 56.4\%; $P < 0.01$), confident in knowing what to ask (81.9\% vs.\ 66.7\%; $P < 0.01$), and confident in making the most of the visit (82.2\% vs.\ 69.2\%; $P < 0.01$), and were more satisfied with the tool (73.9\% vs.\ 61.5\%; $P < 0.05$) and more likely to want to use it again (76.1\% vs.\ 65.4\%; $P < 0.05$).
In the 14-day follow-up survey, visit expectation alignment (78.6\% vs.\ 75.0\%) and agenda topic use (69.0\% vs.\ 66.7\%) were numerically higher with the \agentname{} agents but, as in the main analysis, not statistically significant.

\clearpage
\section{Factor Analysis}
\label{appendix:factor_analysis}

This section supports the factor analysis of patient-reported outcomes (\cref{sec:methods:stats_factors,sec:results:factors}).
We first reproduce the prompt used to extract and categorize the visit expectations and agenda topics covered in each conversation, then define every factor included in the multivariable logistic regression models (\cref{tab:factor_definitions}), and finally report the full model estimates for all five immediate survey items (\cref{fig:supp_factor_analysis}).

\subsection{Visit Expectation and Agenda Topic Categorization}
\label{appendix:prompt_content_themes}

This prompt extracts every distinct item of agent-generated clinical content from the conversation and the pre-visit summary, and sorts it into agenda topics, general health information, and visit expectations. Agenda topics and general health information are coded with a shared, disease-agnostic function taxonomy; visit expectations are coded by the phase of the clinical encounter they refer to. The resulting visit expectation and agenda topic categories are then converted into binary theme indicators for the factor analysis (\cref{appendix:factor_definitions,sec:methods:stats_factors}). In the prompt, \texttt{\{\{convo\}\}} is replaced with the conversation transcript and \texttt{\{\{summary\}\}} with the final pre-visit summary.

The prompt for visit expectation and agenda topic categorization is:
\lstinputlisting[style=prompt]{appendices/prompts/agent_content_themes_prompt.md}

\subsection{Factor Definitions}
\label{appendix:factor_definitions}

\Cref{tab:factor_definitions} lists every factor in the models, its levels and reference category, and how it was defined.

\begin{table}[p]
    \centering
    \scriptsize
    \caption{\textbf{Factors included in the logistic regression models.} Categorical factors are compared with the reference level marked (ref.). Visit expectation and agenda topic themes are binary indicators of whether the conversation or pre-visit summary covered at least one item of that theme, compared with conversations that did not cover it. Visit type, disease area, and the content themes were extracted from the conversation by Gemini 3.6 Flash (\cref{appendix:prompt_visit_type_disease_area,appendix:prompt_content_themes}).}
    \label{tab:factor_definitions}
    \renewcommand{\arraystretch}{1.08}
    \begin{tabularx}{\textwidth}{@{} >{\raggedright\arraybackslash}p{0.17\textwidth} >{\raggedright\arraybackslash}p{0.25\textwidth} >{\raggedright\arraybackslash}X @{}}
        \toprule
        \textbf{Factor} & \textbf{Levels} & \textbf{Definition} \\
        \midrule
        \multicolumn{3}{@{}l}{\textit{Study arm}} \\
        Agent arm & Control (ref.); \Agentname{} & Randomized arm; the \agentname{} arm pools the standard and concise variants. \\
        \addlinespace
        \multicolumn{3}{@{}l}{\textit{Participant characteristics}} \\
        Age group, y & 18--29 (ref.); 30--44; 45--64; 65+ & Participant's age group from the App profile. \\
        Sex at birth & Male (ref.); Female & Participant's sex at birth from the App profile. \\
        \addlinespace
        \multicolumn{3}{@{}l}{\textit{Participant health context}} \\
        Visit type & Specialist consultation (ref.); annual wellness exam; PCP -- chronic condition(s); PCP -- acute condition(s); screening; follow-up / post-procedure; unknown visit type; no visit planned & Type of upcoming visit the participant was preparing for. \\
        Disease area & Cardiovascular \& circulatory (ref.); metabolic \& diabetes management; orthopedic \& musculoskeletal; neurology \& sleep; endocrine \& hormonal health; women's health; dermatology \& skin health; eye, ear \& oral health (HEENT) & Clinical domain of the participant's primary health needs.\\
        \addlinespace
        \multicolumn{3}{@{}l}{\textit{Visit expectations shared by the agent (theme covered vs.\ not)}} \\
        History gathering & Covered; not covered & The clinician will ask about symptoms, history, medications, family or social context. \\
        Physical examination & Covered; not covered & Vitals, physical exam maneuvers, in-office assessments. \\
        Investigations ordered & Covered; not covered & Labs, imaging, or procedures that may be ordered; result turnaround. \\
        Explanation \& diagnostic reasoning & Covered; not covered & The clinician will explain findings, weigh possible causes, or may not have a definitive answer yet. \\
        Treatment planning \& decisions & Covered; not covered & Medication or treatment decisions, options weighed, shared decision-making. \\
        Closure, follow-up \& safety-netting & Covered; not covered & Next appointment, referrals, monitoring plan, when to return or seek help. \\
        Visit logistics \& preparation & Covered; not covered & What to bring, medication list, records, fasting, arrival time, insurance, forms. \\
        \addlinespace
        \multicolumn{3}{@{}l}{\textit{Agenda topics suggested by the agent (theme covered vs.\ not)}} \\
        Cause, etiology \& mechanism & Covered; not covered & Why this is happening; underlying mechanism or pathophysiology; how the affected system works. \\
        Symptom \& diagnosis & Covered; not covered & Characterizing, tracking, or interpreting symptoms, and diagnosis or differential diagnosis. \\
        Test order \& procedure & Covered; not covered & Whether a test or procedure is indicated, how to prepare for it, what it measures, and what results mean. \\
        Treatment options & Covered; not covered & Available treatment or procedure choices, benefits vs.\ risks, trade-offs, watchful waiting. \\
        Medication & Covered; not covered & Dose, side effects, interactions, adherence, refills, starting, stopping, or switching drugs. \\
        Prognosis \& complications & Covered; not covered & Expected course, recovery, recurrence, complications, warning signs, and when to seek urgent care. \\
        Risk, prevention \& screening & Covered; not covered & Personal risk factors, heritability, prevention, screening intervals, immunization. \\
        Self-management & Covered; not covered & Home management, diet, activity, sleep, rehabilitation, symptom relief, self-monitoring. \\
        Care process \& navigation & Covered; not covered & Referrals, care coordination, follow-up scheduling, records, cost or insurance, access to services. \\
        \bottomrule
    \end{tabularx}
\end{table}
\clearpage

\subsection{Detailed Results}
\label{appendix:factor_analysis_full}

\begin{sidewaysfigure}
    \centering
    \includegraphics[width=\textheight,height=0.9\textwidth,keepaspectratio]{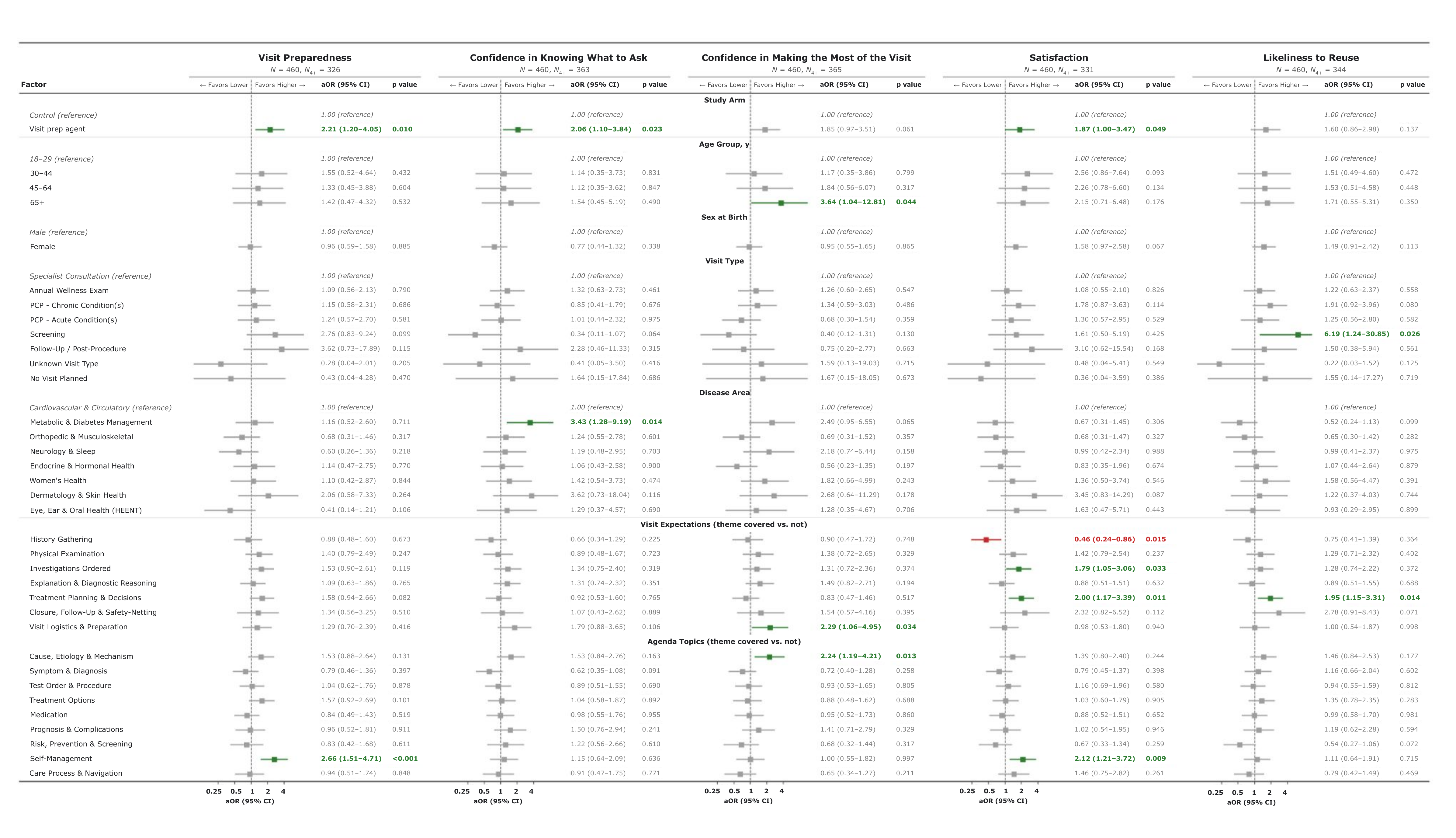}
    \vspace{0.1cm}
    \caption{\textbf{Full Factor Analysis for All Five Immediate Survey Items.} Adjusted odds ratios (aORs, log scale) with 95\% CIs and Wald test $P$ values from separate multivariable logistic regressions of a positive response (rating of 4 or 5) on study arm, age group, sex at birth, visit type, disease area, and the visit expectation and agenda topic themes covered in the conversation. Each row is compared with the reference category in italics (aOR fixed at 1.00), or, for visit expectation and agenda topic themes, with conversations that did not cover the theme. Green and red indicate $P < 0.05$ for higher and lower odds of a positive response, respectively; $P$ values are not adjusted for multiple comparisons. $N$, number of respondents; $N_{4+}$, number of positive responses.}
    \label{fig:supp_factor_analysis}
\end{sidewaysfigure}

\Cref{fig:supp_factor_analysis} reports the adjusted odds ratios for all five immediate survey items; the main text (\cref{fig:factor_analysis}) shows the three visit readiness and confidence items.

% Wide figure, so it is set on its own landscape page.

\clearpage
\section{Joint Analysis of Clinical Ratings and Patient-Reported Outcomes}
\label{appendix:joint_analysis}

This section examines how the clinician and patient perspectives relate to each other for the same conversations (\cref{sec:results:dual_perspective}).
The analyses include sessions that received both a clinical evaluation and an immediate post-interaction survey response.

\subsection{Correlation Between Participant Survey Ratings and Clinical Assessment}
\label{appendix:joint_analysis:correlation}

For each pair of immediate survey item and clinical rating (clinical dimension $\times$ preparation step), we computed Spearman's rank correlation coefficient ($\rho$) with its 95\% confidence interval, separately for each arm and for both arms pooled (\cref{sec:methods:stats_correlation}).
We regarded $\rho > 0.3$ as a moderate correlation and tested whether each correlation exceeded this threshold, controlling the false discovery rate with the BH procedure within each survey item.

\begin{figure}[h!]
    \centering
    \includegraphics[width=\textwidth]{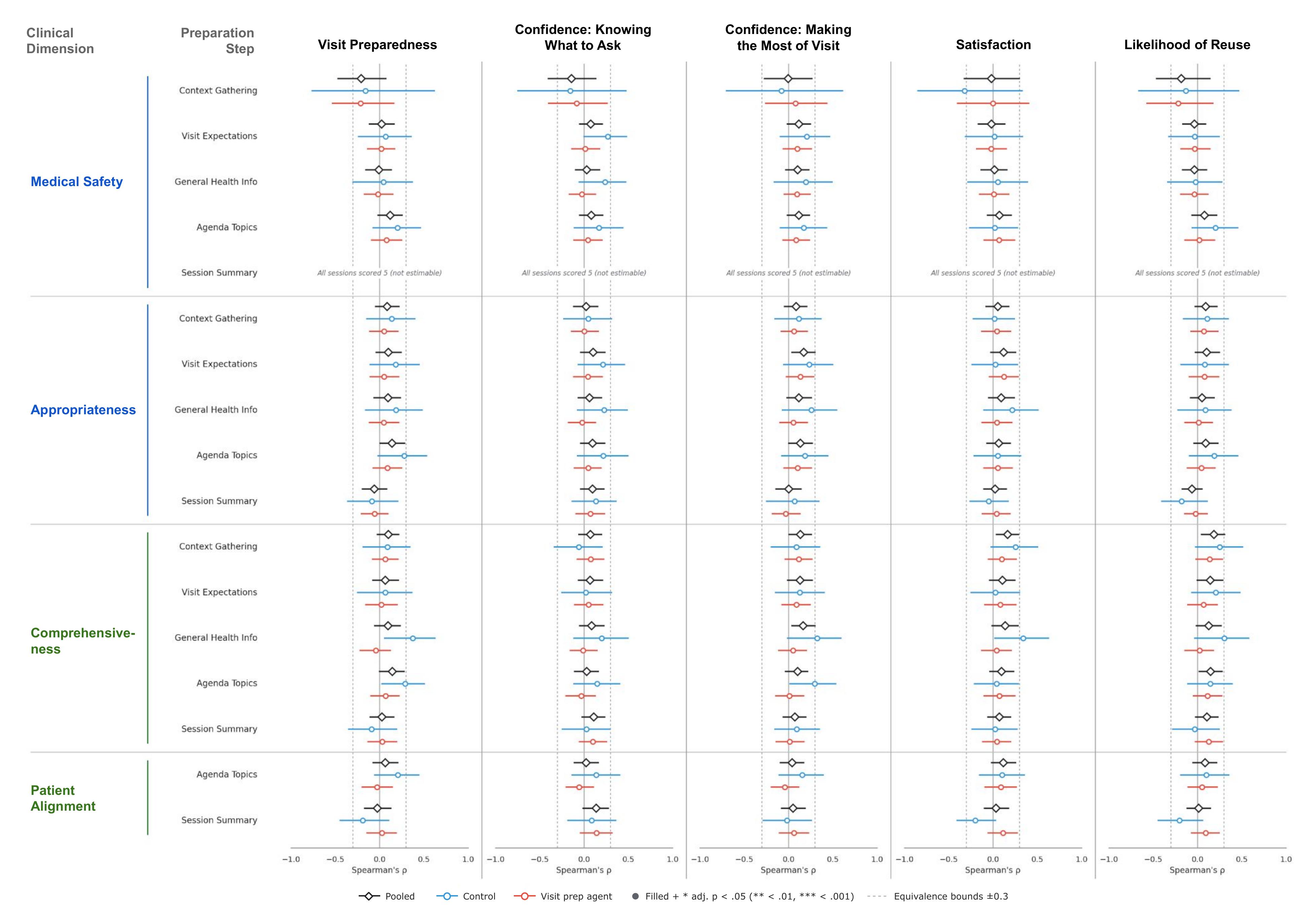}
    \vspace{0.1cm}
    \caption{\textbf{Spearman's Correlation between Participant Survey Responses and Clinician Ratings.} Each row corresponds to the clinical rating of one preparation step on one clinical dimension, and each column to one immediate survey item. Markers show Spearman's $\rho$ for the pooled data (black diamonds), the Control arm (blue), and the \agentname{} arm (red), with 95\% confidence intervals; the \agentname{} arm combines the concise and standard arms. Dashed lines mark $\rho = \pm 0.3$, the threshold for a moderate correlation. Filled markers would indicate a correlation significantly exceeding this threshold after BH correction within each survey item; none reached significance. Medical safety of the pre-visit summary is not estimable because all session summaries were free of safety-related inaccuracies.}
    \label{fig:supp_dual_perspective}
\end{figure}

\Cref{fig:supp_dual_perspective} shows the resulting coefficients.
Correlations were weak throughout: nearly all point estimates fell within $\pm 0.3$, and after correction no pair of survey item and clinical rating showed a significant correlation in either arm or in the pooled data.
The correlation for medical safety of the pre-visit summary could not be estimated because every session received a rating of 5.
The intervals of the Control arm are wider than those of the \agentname{} arm, reflecting its smaller number of sessions.

\newpage
\setlength\bibitemsep{3pt}
\printbibliography
\clearpage
\end{refsection}

\end{document}